\documentclass[sigconf]{acmart}
\usepackage{hyperref} 
\usepackage{graphicx}
\usepackage{amsmath}
\usepackage{amsthm}
\usepackage{amsfonts}
\usepackage{subfigure}
\usepackage{balance}
\usepackage[ruled,vlined]{algorithm2e}
\usepackage{multirow}
\usepackage{enumitem}
\usepackage{threeparttable} 

\usepackage{multirow}
\usepackage{makecell}
\usepackage{pifont}% http://ctan.org/pkg/pifont
\newcommand{\cmark}{\ding{51}}%
\newcommand{\xmark}{\ding{55}}%

\newtheorem{exam}{Example}

\newcommand{\squishlist}{
 \begin{list}{$\bullet$}
  {  \setlength{\itemsep}{0pt}
     \setlength{\parsep}{3pt}
     \setlength{\topsep}{3pt}
     \setlength{\partopsep}{0pt}
     \setlength{\leftmargin}{2em}
     \setlength{\labelwidth}{1.5em}
     \setlength{\labelsep}{0.5em}
} }
\newcommand{\squishlisttight}{
 \begin{list}{$\bullet$}
  { \setlength{\itemsep}{0pt}
    \setlength{\parsep}{0pt}
    \setlength{\topsep}{0pt}
    \setlength{\partopsep}{0pt}
    \setlength{\leftmargin}{2em}
    \setlength{\labelwidth}{1.5em}
    \setlength{\labelsep}{0.5em}
} }

\newcommand{\squishdesc}{
 \begin{list}{}
  {  \setlength{\itemsep}{0pt}
     \setlength{\parsep}{3pt}
     \setlength{\topsep}{3pt}
     \setlength{\partopsep}{0pt}
     \setlength{\leftmargin}{1em}
     \setlength{\labelwidth}{1.5em}
     \setlength{\labelsep}{0.5em}
} }

\newcommand{\squishend}{
  \end{list}
}

\newcommand{\eat}[1]{}

\newcommand{\kw}[1]{{\ensuremath {\mathsf{#1}}}\xspace}

\newcommand{\stitle}[1]{\vspace{1.5ex}\noindent{\bf #1}}
\newcommand{\etitle}[1]{\vspace{0.8ex}\noindent{\em #1}}
\newcommand{\eetitle}[1]{\vspace{0.8ex}\noindent{\em\underline{#1}}}

\newcounter{ccc}

\newcommand\redout{\bgroup\markoverwith
{\textcolor{red}{\rule[.5ex]{2pt}{2pt}}}\ULon}

\newcommand{\mips}{\kw{MIPS}}
\newcommand{\anns}{\kw{ANNS}}

\newcommand{\vess}{\kw{VSS}}

\newcommand{\iceberg}{\kw{Iceberg}}
\newcommand{\rag}{\kw{RAG}}
\newcommand{\nns}{\kw{NNS}}

\newcommand{\nn}{\kw{NN}}
\newcommand{\ip}{\kw{IP}}
\newcommand{\magg}{\kw{MAG}}
\newcommand{\hnsw}{\kw{HNSW}}

\newcommand{\scann}{\kw{ScaNN}}

\newcommand{\graphtype}{\small Graph}      
\newcommand{\parttype}{\small Partition}

\usepackage[tikz]{bclogo}

\AtBeginDocument{%
  }

\setcopyright{acmlicensed}
\copyrightyear{2026}
\acmYear{2026}
\acmDOI{XXXXXXX.XXXXXXX}
\acmConference[SIGMOD'26]{ACM International Conference on Management of Data}{2026}{Bengaluru, India}
\acmISBN{978-1-4503-XXXX-X/2018/06}
\setcopyright{none}
\renewcommand\footnotetextcopyrightpermission[1]{}
\begin{document}

\title{Reveal Hidden Pitfalls and Navigate Next Generation of Vector Similarity Search from Task-Centric Views}

\author{Tingyang Chen}
\orcid{0009-0008-5635-9326}
\affiliation{%
  \institution{Zhejiang University}
  \city{Hangzhou}
  \country{China}
}
\email{chenty@zju.edu.cn}

\author{Cong Fu}
\orcid{0000-0002-3624-6665}
\affiliation{%
  \institution{Shopee Pte. Ltd.}
  \city{Singapore}
  \country{Singapore}}
\email{fc731097343@gmail.com}

\author{Jiahua Wu}
% \orcid{0000-0002-3624-6665}
\affiliation{%
  \institution{Shopee Pte. Ltd.}
  \city{Singapore}
  \country{Singapore}}
\email{gauvain.wujiahua@gmail.com}

\author{Haotian Wu}
% \orcid{0009-0008-5635-9326}
\affiliation{%
  \institution{Zhejiang University}
  \city{Hangzhou}
  \country{China}
}
\email{haotian.wu@zju.edu.cn}

\author{Hua Fan}
% \orcid{0009-0008-5635-9326}
\affiliation{%
  \institution{Alibaba Cloud Computing}
  \city{Hangzhou}
  \country{China}
}
\email{guanming.fh@alibaba-inc.com}

\author{Xiangyu Ke}
\orcid{0000-0001-8082-7398}
\authornote{Corresponding author.}

\affiliation{%
  \institution{Zhejiang University}
  \city{Hangzhou}
  \country{China}
}
\email{xiangyu.ke@zju.edu.cn}

\author{Yunjun Gao}

\orcid{0000-0003-3816-8450}
\affiliation{%
  \institution{Zhejiang University}
  \city{Hangzhou}
  \country{China}
}
\email{gaoyj@zju.edu.cn}

\author{Yabo Ni}
\orcid{0000-0002-7535-8125}
\affiliation{%
  \institution{Nanyang Technological University}
  \city{Singapore}
  \country{Singapore}}
\email{yabo001@e.ntu.edu.sg}

\author{Anxiang Zeng}
\orcid{0000-0003-3869-5357}
\affiliation{%
  \institution{Nanyang Technological University}
  \city{Singapore}
  \country{Singapore}}
\email{zeng0118@ntu.edu.sg}


\begin{abstract}

Vector Similarity Search (\vess) in high-dimensional spaces is rapidly emerging as core functionality in next-generation database systems for numerous data‑intensive services -- from embedding lookups in large language models (\textsf{LLMs}), to semantic information retrieval and recommendation engines.
Current benchmarks, however, evaluate \vess primarily on the recall–latency trade‑off against a ground truth defined solely by distance metrics, neglecting how retrieval quality ultimately impacts downstream tasks. This disconnect can mislead both academic research and industrial practice.

We present \iceberg, a holistic benchmark suite for end‑to‑end evaluation of \vess methods in realistic application contexts.
From a task-centric view, \iceberg uncovers the {\bf Information Loss Funnel}, which identifies three principal sources of end‑to‑end performance degradation: 
(1) {\em Embedding Loss} during feature extraction; (2) {\em Metric Misuse}, where distances poorly reflect task relevance; (3) {\em Data Distribution Sensitivity}, highlighting index robustness across skews and modalities. For a more comprehensive assessment, \iceberg spans eight diverse datasets across key domains such as image classification, face recognition, text retrieval, and recommendation systems. Each dataset, ranging from 1M to 100M vectors, includes rich, task‑specific labels and evaluation metrics, enabling assessment of retrieval algorithms within the full application pipeline rather than in isolation. 
\iceberg benchmarks 13 state-of-the-art \vess methods and re-ranks them based on application-level metrics, revealing substantial deviations from traditional rankings derived purely from recall-latency evaluations. Building on these insights, we define a set of task-centric meta-features and derive an interpretable decision tree to guide practitioners in selecting and tuning \vess methods for their specific workloads. 
 Finally, we outline promising avenues for future research aimed at closing the gap between synthetic \vess benchmarks and production deployments.

\end{abstract}

\maketitle
\vspace{-1.5mm}
\section{Introduction}
\label{sec:intro}
\vspace{-0.5mm}

\begin{figure}[tb!]
\vspace{0.5ex}
\centering
\centerline{\includegraphics[width=1\linewidth]{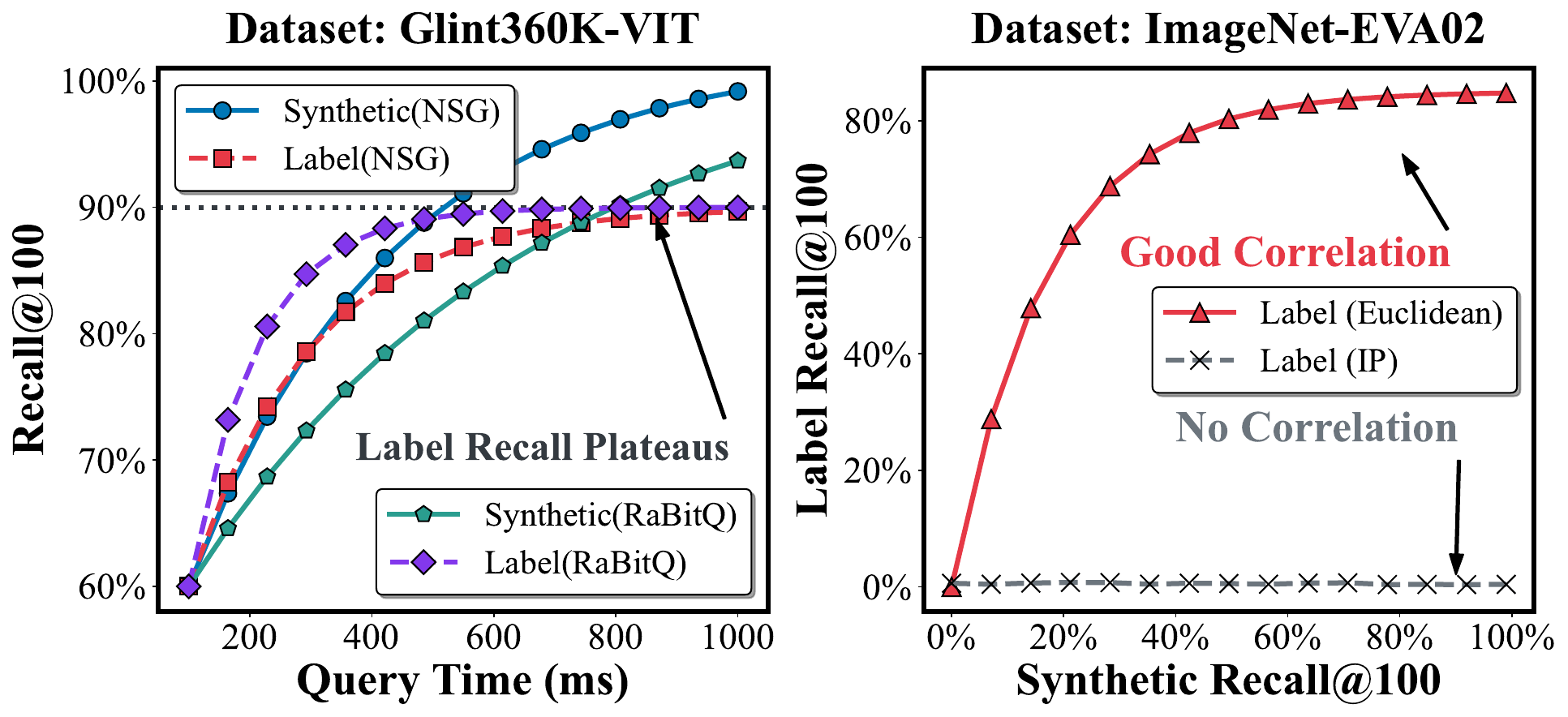}}
\vspace{-3ex}
\caption{Illustration of the gap between synthetic recall and task-related performance. Left: face recognition dataset; right: image classification dataset. See \S\ref{sec:dataset} for dataset details.}
\vspace{-3.5ex}
\label{fig:example_1}
\end{figure}

\begin{figure*}[tb!]
\vspace{1ex}
\centering
\centerline{\includegraphics[width=1\linewidth]{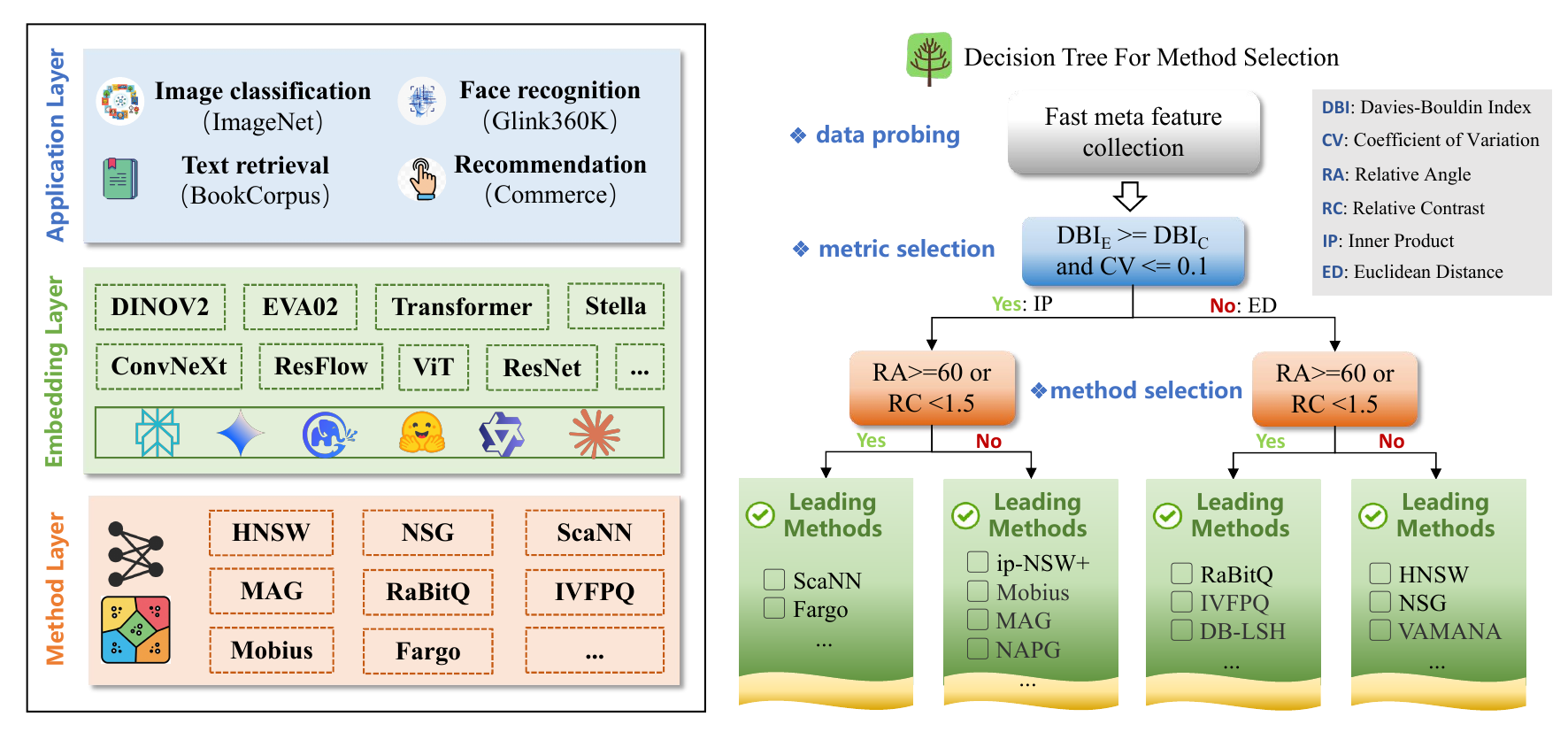}}
\vspace{-2ex}
\caption{The main contribution of \iceberg. Detailed information on the datasets and methods is provided in \S\ref{sec:dataset}, while the construction of the decision tree and the computation of meta-features are described in \S\ref{sec:exp}.}
\vspace{-1ex}
\label{fig:overview}
\end{figure*}
{\em Vector Similarity Search} (\vess) in high-dimensional spaces serves as a core primitive in modern databases~\cite{pan2024survey} for various applications like recommendation systems~\cite{gao2018recommendation} and retrieval-augmented generation (\rag) pipelines~\cite{lewis2020retrieval}. 
As adoption widens, practitioners must choose \vess methods that satisfy both throughput and \textbf{task-level effectiveness}. 
Existing benchmarks~\cite{aumuller2020ann, li2019approximate, kangbigvectorbench} primarily measure the recall-speed trade-off against ground truths derived solely from distance metrics (e.g., Euclidean or Inner-Product). 
These \textbf{synthetic metrics} abstract away from task semantics, ignoring application-specific labels or utility, and thereby obscure how retrieval quality propagates to downstream outcomes, leading to a disconnect between academic evaluations and practical deployment criteria.
In our preliminary analysis from a task-centric view, we identify a long-overlooked pitfall in \vess, termed the \textbf{Information Loss Funnel}, which helps explain why high-synthetic recall retrieval does not necessarily guarantee strong end-task performance, as illustrated below:

\begin{exam}
Figure~\ref{fig:example_1} contrasts a widely used synthetic metric (synthetic recall@100) with a downstream-aware measure (\textit{label recall}), on two public datasets. Synthetic Recall@100 evaluates how close the retrieved vectors to a query in terms of distance (e.g., Euclidean or Inner Product), whereas label recall measures the fraction of those 100 retrieved items whose class or identity label matches the query. 

On \textsf{Glint360K-ViT} for face recognition (left), \textsf{NSG}~\cite{fu2019fast} attains 99\% synthetic recall@100 faster than \textsf{RaBitQ}~\cite{gao2024rabitq}, yet achieves only 90\% label recall -- and does so more slowly than \textsf{RaBitQ}. This divergence reveals that superior synthetic recall need not translate to better task effectiveness. 
The ImageNet classification case (right) is even more dramatic: with Euclidean distance, synthetic recall@100 and label recall track closely, but under inner‑product distance, synthetic recall@100 soars to 99.9\% while label recall collapses below 1\%. In other words, the index returns almost exclusively wrong‐class images despite an ostensibly ideal benchmark score, demonstrating that conventional recall–latency evaluations can be entirely misleading for real‑world tasks with a mismatched metric.
\end{exam}

These findings demonstrate that optimizing synthetic recall metrics alone neither ensures improved task‑level performance nor reveals the most appropriate similarity measure. 
We trace the mismatch to a three-stage  \textit{Information Loss Funnel}: 
\textbf{Layer 1. Embedding Loss}: Mapping raw data to fixed-length vectors inevitably discards fine-grained information. Hence, vector proximity does not always align with task-specific utility.  
\textbf{Layer 2. Metric Misuse}: 
Applying a generic distance measure that poorly matches the geometry of the learned embeddings can amplify errors, especially when the embedding model is trained with complex objectives \cite{fu2024residual,huang2022riemannian} beyond pure metric learning \cite{kulis2013metric,hsieh2017collaborative}. 
\textbf{Layer 3: Distribution Sensitivity}: \vess methods differ in how they balance recall vs. latency under varying data distributions. Choosing an index that lacks robustness to the specific vector distribution in use can incur significant degradation in both throughput and end‑task accuracy.

\begin{table*}[tb!]
    \centering
    %\vspace{-2mm}
     \caption{Comparison of \iceberg with state-of-the-art vector similarity search benchmarks.
    "Scale" denotes the size of the benchmark datasets.
    "Embedding Model" indicates the advancement level of the models employed by the benchmark.
    "Funnel Scope" indicates whether the benchmark provides a comprehensive analysis of the information loss funnel.
    "Domain" specifies which similarity search problem the benchmark focuses on.
    "Downstream Utility" indicates whether the benchmark considers the effectiveness of downstream tasks.
    "Explainability" refers to whether the method rankings provided by the benchmark are interpretable for users to select appropriate methods. }
    \label{tab:benchmark-analysis}
    \vspace{-1mm}
 \resizebox{\linewidth}{!}{
    \begin{tabular}{c c c c c c c c c c}
            \textbf{Benchmark Name} & \textsc{Scale} & \textsc{Embedding Model} &  \textsc{Funnel Scope} & \textsc{Domain} & \textsc{Downstream Utility} & \textsc{Explainability}\\ 
            \midrule
		  {\bf ANN-Benchmark}~\cite{aumuller2020ann} & Small & Out-dated & Partial & \anns & \xmark & \xmark\\ 
            {\bf Big-ANN-Benchmarks}~\cite{simhadri2022results} & Large & Modest & Partial & \anns & \xmark & \xmark\\
            {\bf BigVectorBench}~\cite{kangbigvectorbench} & Large  & New & Partial  &  \anns& \xmark & \xmark\\ 
            {\bf MTEB}~\cite{muennighoff2023mteb} & Small  & New & Partial  &  \anns & \xmark & \xmark \\ 
            \midrule
            {\bf IceBerg (this work)} & Large & New & Full & \anns \& \mips & \cmark & \cmark\\
            %\bottomrule
    \end{tabular}}
    \vspace{0ex}
    \label{comparison}
\end{table*}

However, most public benchmarks probe only the third layer, leaving the funnel’s upper stages untested. To remedy this oversight, we introduce \textbf{Iceberg}, a holistic, task-centric benchmark suite that exposes the hidden pitfalls of \vess evaluation and guides future algorithmic development. \textsf{Iceberg} makes a continuous and tightly integrated set of contributions. It first introduces the information-loss funnel, which exposes critical yet overlooked issues in the existing literature. It then provides a suite of benchmarks with newly labeled datasets that empirically validate this model. Building on these analyses, it further proposes a selection method that operationalizes the framework’s insights, enabling end users to make task-aligned algorithmic choices. Finally, it highlights promising directions for bridging the gap between academic research and industrial practice. The main contributions of \iceberg are summarized as follows and visualized in Figure~\ref{fig:overview}:

\stitle{(1) Information Loss Funnel: A Fine‑Grained Diagnostic Model.} 
We introduce the Information Loss Funnel, a three‑layer framework that systematically diagnoses why vector similarity search methods fail to deliver on task‑level objectives. 
By in-depth examining (1) Embedding Loss, (2) Metric Misuse, and (3) Distribution Sensitivity, our model unifies disparate research threads under a common taxonomy. 
This layered perspective not only reveals hidden failure modes masked by synthetic benchmarks but also guides researchers toward the most impactful optimization targets.

\stitle{(2) Comprehensive Task-centric Evaluation Benchmark.} To assess and validate the diagnostic model, we release  a broad benchmark suite spanning eight diverse datasets from four representative end-application domains -- image classification, face recognition, text retrieval, and large-scale recommendation -- each with 1M-100M vectors generated from state-of-the-art embedding models~\cite{dinov2_base,vit_tiny_patch8_112,stella_en_1_5B,resflow_github} and paired with task-specific ground truths.
We evaluate 13 leading \vess methods (covering tree-, hashing-, clustering-, and graph-based) on both Approximate Nearest Neighbor Search (\anns) \cite{ram2012maximum,koenigstein2012efficient,gao2024rabitq,aguerrebere2023similarity,datar2004locality,malkov2018efficient,fu2019fast} and Maximum Inner Product Search (\mips) \cite{chen2025maximum,chen2025stitching,zhao2023fargo,morozov2018non,guo2020accelerating,ma2024reconsider}. \textbf{Iceberg} reports not only synthetic recall and throughput but also downstream utility measured by label recall@K, hit@K, and matching score@K (detailed in \S\ref{sec:dataset}). Our results reveal substantial re-rankings compared to the traditional distance-only leaderboards, offering a more faithful guide for real-world deployments. Moreover, we develop an extensible pipeline that facilitates the integration of future tasks.

\stitle{(3) Explainable Decision Tree for Method Selection.} 
Leveraging the Information Loss Funnel model and extensive experimental analyses on benchmark datasets, we develop a lightweight, transparent decision‑tree toolkit that maps dataset characteristics to suitable \vess methods. 
We identify easy-to-compute meta‑features -- such as {\em clustering tightness}, {\em vector norm distributions}, {\em angular separations}, and {\em distance contrast} -- that correlate strongly with end‑task performance. 
They form the decision nodes of our tree, enabling practitioners to rapidly extract them from {\em any} dataset and traverse the tree to determine the optimal \vess indexing and search strategy. 
To the best of our knowledge, this is the {\em first} public toolkit linking data traits directly to \vess method choices 
in a task‑centric manner.

\stitle{(4) Insights of Future Direction.} 
Analysis of the \textbf{Information Loss Funnel} highlights three promising paths to bridge the gap between benchmarks and applications: 
\textbf{(1)} \textit{Task-Aware} \vess: Jointly optimizes indexing and searching strategies with downstream objectives in mind; 
\textbf{(2)} \textit{Metric-Aware} \vess: 
Automatically select, combine, or desensitize similarity metrics to accommodate embedding learned under complex, non-metric losses;  
and \textbf{(3)} \textit{Distribution-Aware} \vess: Dynamically adapt the index structures and retrieval strategies to heterogeneous data distribution, e.g., skewed densities or segmented user behavior, to sustain both accuracy and efficiency.

\stitle{Roadmap.} \S\ref{sec:background} reviews previous related works and background. \S\ref{sec:dataset} provides a detailed description of the datasets, end tasks, methods, and overviews the design philosophy behind \iceberg. \S\ref{sec:exp} raises the
research questions, then provides a comprehensive analysis based
on the experimental outcomes of \iceberg, and outlines the future directions. \S\ref{sec:discuss} discusses the implications and limitations of \iceberg.
\S\ref{sec:conclude} concludes this paper.

%\vspace{-1ex}
\section{Background}
\label{sec:background}

\subsection{The Background of Vector Similarity Search}
%\vspace{-1ex}
Vector similarity search encompasses two core problems, Approximate Nearest Neighbor Search (\anns) in Euclidean space and Maximum Inner Product Search (\mips) in inner‑product space, and has spawned a rich ecosystem of methodic approaches. These methods fall into four principal categories based on their approaches and methodologies to data processing: 
(1) Tree-based methods~\cite{ram2012maximum,koenigstein2012efficient,ram2019revisiting,ma2024reconsider} recursively partition the vector space into hierarchical cells, offering fast query pruning but suffering as dimensionality increases; 
(2) Hash-based methods~\cite{datar2004locality,huang2015query,liu2014sk,zhao2023fargo,shrivastava2014asymmetric,wei2024det,tian2023db} map similar vectors into the same hash buckets, enabling sublinear search at the cost of controlled collision errors; 
(3) Clustering methods~\cite{gao2024rabitq,aguerrebere2023similarity,guo2020accelerating,matsui2018survey,babenko2014additive,ferhatosmanoglu2000vector,gao2025practical} compress the dataset into representative centroids or quantized codes, reducing both storage and computation but introducing quantization error;
%, which reduce both computational cost and memory consumption by adopting various vector compression techniques; 
and (4) Graph-based methods~\cite{fu2019fast,malkov2018efficient,morozov2018non,chen2025stitching, chen2025maximum, zhou2019mobius,peng2023efficient,chen2024roargraph,lu2021hvs}, which construct complicated proximity graphs that capture neighborhood relationships well in high-dimensional space, thus turning search into a fast graph traversal with strong recall–latency trade‑offs.
For a comprehensive review, we direct the readers to recent tutorials and surveys~\cite{wang2021comprehensive,pan2024survey,matsui2018survey,azizi2025graph}. In traditional benchmarks, synthetic recall is commonly employed to evaluate the accuracy of \textsf{VSS}. It is commonly defined as the intersection between the retrieved top-k vectors and the ground-truth top-k neighbors.

\subsection{The Current Popular Benchmarks}

As vector similarity search (\vess) matures, standardized benchmarks are essential to assess vector quality, algorithmic advances and guide practical adoption in real-world deployment. 
Prominent suites such as \textsf{ANN}-Benchmarks\cite{aumuller2020ann}, Big-\textsf{ANN}-Benchmarks\cite{simhadri2022results},  BigVectorBench~\cite{kangbigvectorbench}, and \textsf{MTEB}~\cite{muennighoff2023mteb} have made valuable contributions, but they mainly emphasize synthetic recall under fixed metrics and offer limited attention for end-to-end evaluation.
We provide a structured comparison in Table~\ref{tab:benchmark-analysis}, along six key dimensions: dataset scale, embedding model, funnel coverage, metric domain, downstream utility, and explainability.

First, in terms of dataset scale and embedding quality, \textsf{ANN}-Benchmarks is constrained to relatively small, legacy datasets (e.g., \textsf{SIFT1M}, \textsf{GLOVE1M}) derived from outdated embedding models (e.g., Eigen decomposition, \textsf{GoogLeNet}).
As a result, these datasets fail to capture the complexity and semantic richness of modern applications. 
Big-ANN-Benchmarks and BigVectorBench represent progress by adopting larger-scale datasets and incorporating more recent embedding techniques. \textsf{MTEB} incorporates more advanced embedding models—including \textsf{LLM}-based ones—and focuses on evaluating the representation quality of different embeddings. However, its dataset scale is relatively small. Second, regarding evaluation scope—specifically funnel coverage and metric domain, all four benchmarks focus almost exclusively on \anns and assess only limited portions of the information-loss funnel. \textsf{MTEB} is limited to Layer 1 (Embedding Loss), contrasting vectors from different representation models on a common task. It normalizes by default and disregards similarity-metric selection and distribution-dependent retrieval variance. The remaining benchmarks probe only Layer 3 (Distribution Sensitivity), neglecting the impacts derived from other layers.
Moreover, they offer no evaluation for \mips, despite its prevalence in recommendation and ranking systems. Third, existing benchmarks provide minimal or no support for task‑level utility analysis (e.g., classification accuracy, recommendation hit‑rate) and lack tools to explain why one method outperforms another in a given downstream context. Consequently, these benchmarks center on synthetic recall under a chosen metric. While this contributes to evaluation in some respects, users gain little actionable guidance for selecting methods aligned with their application objectives.

\iceberg addresses these deficiencies by spanning both \anns and \mips, employing large‑scale, modern embeddings, and integrating end‑to‑end task utility assessment. 
Crucially, it delivers full-funnel coverage -- including embedding fidelity and metric alignment probes -- and an explainable decision tree that maps dataset meta-features to recommended methods. This holistic approach empowers practitioners to diagnose failure modes, interpret performance trade‑offs, and make informed choices tailored to their specific workloads.

\subsection{Mainstream Task-Centric Metrics for VSS}

Vector similarity Search (\vess) underlines many real-world systems. In this work, \iceberg examines \vess in four important application domains: \textit{image classification}, \textit{face recognition}, \textit{text retrieval}, and \textit{recommendation systems}.

In these practical applications, industrial practitioners are not only concerned with the trade-off between synthetic recall and query throughput (QPS), but also prioritize realistic downstream task-centric metrics that reflect the true utility of the \vess system. Different applications have different goals for downstream tasks from vector similarity search. 
For image classification, the main concern is whether the returned images reflect the true query class, which is in favor of large scale image labeling and recognition. 
In face recognition, an important downstream task is to recall identity vectors corresponding to the query, prioritizing accuracy for security and payment. 
In text retrieval, operations involve determining whether a query can recall semantically relevant passages, measuring content matching accuracy. 
Recommendation system metrics assess if item embeddings reflect users' preference for items and if they emphasize high-relevance, popular items like bestsellers, for the purpose of conversion and personalization. 
Despite the ubiquity of synthetic benchmarks (e.g., synthetic recall@K), these metrics can diverge sharply from task objectives -- a mismatch we attribute to the Information Loss Funnel. By evaluating \vess methods against both synthetic and realistic task‑centric metrics across diverse applications, \iceberg reveals where and why high synthetic recall may fail to deliver real‑world effectiveness.

\section{Resource Collection for Funnel Analysis}
\label{sec:dataset}

\begin{figure}[tb!]
\vspace{1ex}
\centering
\centerline{\includegraphics[width=1\linewidth]{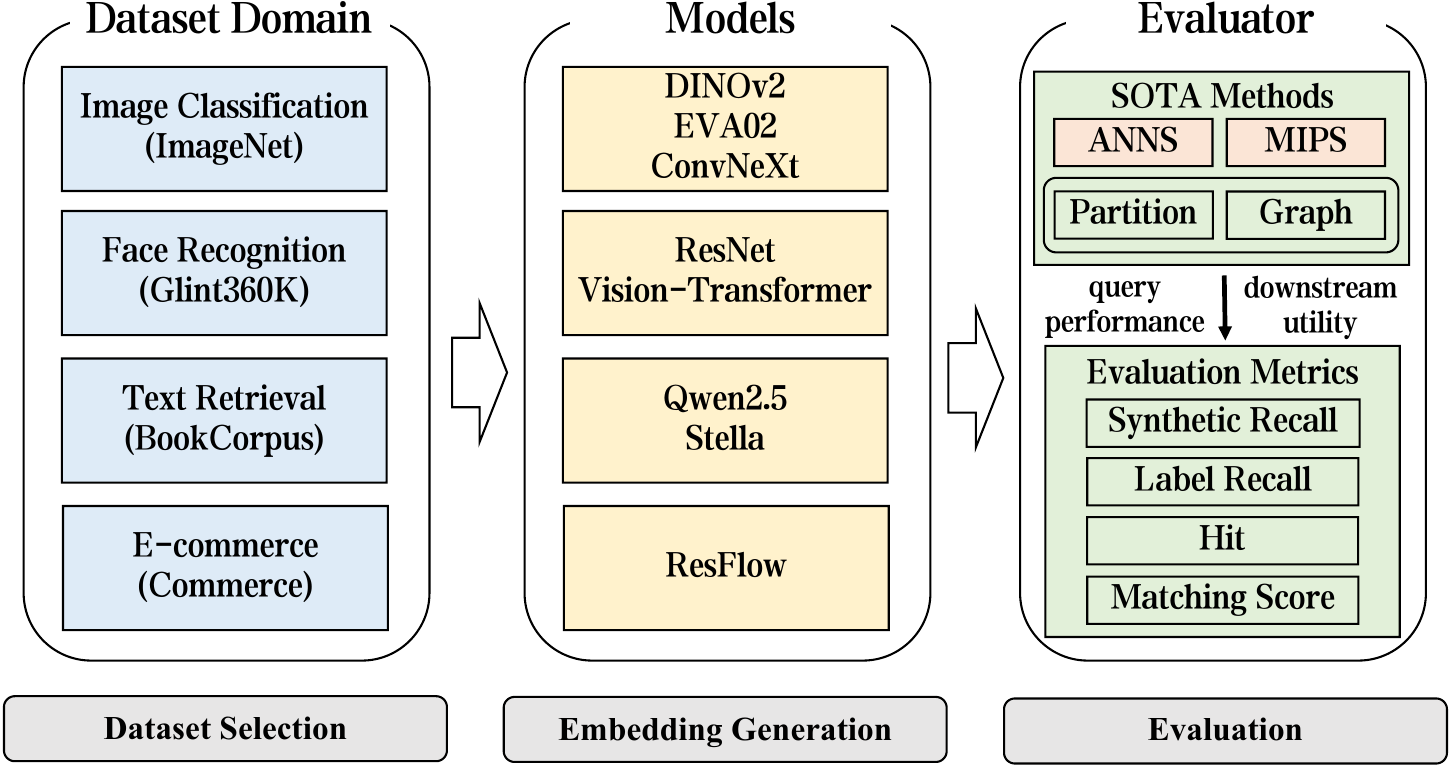}}
\vspace{-2ex}
\caption{Benchmark Pipeline.}
\vspace{-2ex}
\label{fig:pipeline}
\end{figure}

To operationalize the information loss funnel, \iceberg spans four representative, real‑world tasks—image classification, face recognition, text retrieval, and recommendation. 
In \S~\ref{sec:data_process}, we describe our data processing pipeline, including dataset selection criteria, task abstraction, and embedding generation for each application. 
\S~\ref{sec:algorithm_evaluation} presents the method evaluation setup, detailing the 13 state‑of‑the‑art \vess methods, their configurations, theoretical properties, and performance metrics. 
In \S~\ref{sec:reproducibility}, we outline our reproducibility framework to enable longitudinal studies of the information loss funnel. 
Figure~\ref{fig:pipeline} illustrates the end‑to‑end \iceberg workflow, and Table~\ref{tab:prop} summarizes the domains, dataset scales, embedding models, and downstream metrics.

\begin{table*}[tb!]
  \caption{Overview of benchmark datasets}
  \label{tab:prop}
  \resizebox{0.99\linewidth}{!}{
  \begin{tabular}{cccccccc}
    \hline
    \textbf{Task} & \textbf{Dataset} & \textbf{Data Num}  & \textbf{Dim} & \textbf{Embedding Model} & \textbf{Task-specific Label} & \textbf{Task-centric Metrics} \\
    \hline

    \multirow{4}{*}{\makecell[c]{\textbf{Image}\\\textbf{Classification}}} 

    & \textsf{ImageNet-DINOv2} & 1,281,167 &  768 &  \textsf{DINOv2}~\cite{oquab2024dinov2} & Image Class & Label Recall@K \\
    & \textsf{ImageNet-EVA02} & 1,281,167 &  1024 & \textsf{EVA02}~\cite{fang2024eva} &  Image Class  & Label Recall@k \\
    & \textsf{ImageNet-ConvNeXt} & 1,281,167 &  1536 & \textsf{ConvNeXt}~\cite{woo2023convnext} &  Image Class  & Label Recall@K\\
     & \textsf{ImageNet-AlexNet} & 1,281,167 &  4096 &  \textsf{AlexNet}~\cite{krizhevsky2012imagenet} & Image Class & Label Recall@K \\
    \hline

    \multirow{2}{*}{\makecell[c]{\textbf{Face}\\\textbf{Recognition}}} 
    & \textsf{Glint360K-ViT} & 17,091,649 & 512 & \textsf{ArcFace-IR101}~\cite{deng2019arcface} & Identity ID  & Label Recall@K\\
    & \textsf{Glint360K-IR101} & 17,091,649 &  512 & \textsf{ArcFace-ViT}~\cite{deng2019arcface} &  Identity ID  & Label Recall@K \\
    \hline

    \makecell[c]{\textbf{Text}\\\textbf{Retrieval}} 
    & \textsf{BookCorpus} & 9,250,529 & 1024 & \textsf{Stella\_en\_1.5B\_v5} &  Paragraph ID  & Hit@K\\
    \hline

    \makecell[c]{\textbf{Recommendation}}
    & \textsf{Commerce} & 99,085,171 &  48 & \textsf{ResFlow}~\cite{fu2024residual} & Item ID & Matching Score@K\\
    \hline
  \end{tabular}}
\end{table*}

\subsection{Datasets for Important Application}
\label{sec:data_process}
\stitle{(1) Image Classification.} Image classification serves as a canonical use case for vector‑based retrieval, aiming to align the knowledge representation of images with human conceptual understanding. The goal of \vess is to retrieve visually and semantically similar images for downstream tasks that align with human perception. 

\eetitle{Data Collection.} We use the \textsf{ImageNet} dataset~\cite{imagenet_challenge} as the representative benchmark for image classification. This large-scale dataset contains 1,000 object categories, with 1,281,167 training images and 50,000 validation images, each annotated with ground-truth labels. Due to its broad coverage and widespread adoption in the computer vision community, \textsf{ImageNet} remains a standard resource for training and evaluating image classification models.

\eetitle{End Task Design.} The image category annotations are treated as task-specific labels. An essential criterion for assessing whether the retrieved vectors satisfy the needs of the downstream task is the consistency between the labels of the retrieved images and that of the query. Thus, we define Label Recall@K as a task-centric metric.

\etitle{\textbf{Label Recall@K}.} It measures how many correct task-specific labels appear in the top-K retrieved results. It captures the fraction of retrieved vectors whose task-relevant labels match the query’s label, providing a practical measure of downstream task performance.
\begin{equation}
\label{eq:label_recall}
    \text{Label Recall}@K = \frac{1}{|\mathcal{Q}|*K} \sum_{\mathbf{q} \in \mathcal{Q}} \sum_{i=1}^{K}\mathcal{I}(L(R_i)=L(q))
\end{equation}
where $\mathcal{Q}$ is the set of queries, $L(x)$ is a label function to get label for vector $x$, $\mathcal{I}(\cdot)$ is an identity function returning 1 if input statement holds and 0 otherwise, and $R_i$ means the $i$th vector in the retrieved vector set $R$ with size $K$.

\eetitle{Embedding Generation.} Image embedding models are rapidly advancing with diverse representation strategies. To compare the impact of different models on the embedding space and downstream tasks, we evaluate representative state-of-the-art (leading methods in terms of classification accuracy under different research venues) backbone models: \textsf{DINOv2}~\cite{dinov2_base}, \textsf{EVA02}~\cite{eva02_large_patch14}, \textsf{ConvNeXt}~\cite{convnext_large_mlp}. Furthermore, we used an old embedding model, AlexNet~\cite{krizhevsky2012imagenet}. We did not include it in the major experiments, but rather used it to compare old and modern embeddings, illustrating how representation shifts affect downstream performance.

\stitle{(2) Face Recognition.} Face recognition is a specialized and highly important visual retrieval task, with widespread applications in areas such as security and payment. Unlike general visual classification, it demands extremely fine-grained accuracy to differentiate between individuals. The goal of \vess is to retrieve vectors corresponding to the same identity as the query face.

\eetitle{Data Collection.} We adopt the open-source \textsf{Glint360K} dataset~\cite{an2021partial} as the representative benchmark for face recognition. This large-scale dataset contains over 17 million images spanning 360,232 identities, curated from public sources and annotated via a combination of manual and automated methods. For query construction in \iceberg, we randomly sample one face image per identity, with the first 20,000 identities used as the test set.

\eetitle{End Task Design.} The identity ID corresponding to each face embedding is used as the task-specific label. Although the granularity of the tasks differs, face recognition and image classification share the same task-centric evaluation metric: Label Recall@K (refer to Equation~\ref{eq:label_recall}). This metric effectively captures the degree of identity match and serves as a reliable task-centric measure.

\eetitle{Embedding Generation.} To emphasize the architectural diversity of the models, we selected representative backbones from both the \textsf{CNN}~\cite{cvlface_arcface_ir101} and \textsf{ViT}~\cite{vit_tiny_patch8_112} paradigms, and employed \textsf{Arcface}, a leading method specifically designed for face recognition, during training.

\stitle{(3) Text Retrieval.} Text retrieval is a critical problem in \vess, especially in the era of \textsf{LLMs}, where it plays a central role in \rag frameworks to mitigate hallucinations commonly observed in generative models. In text retrieval, the goal of \vess is to retrieve the text passages that best match a given query sentence, serving as an external knowledge source to provide relevant information.

\eetitle{Data Collection.} We use a cleaned version of the open-source \textsf{BookCorpus} dataset, which contains text from approximately 19,000 books across diverse domains. The corpus was segmented into 9,250,529 paragraph-level chunks, each formed by concatenating eight sentences. From this collection, 10,000 paragraphs were randomly selected to extract the query set.

\eetitle{End Task Design.} The unique ID of each paragraph was used as the label. The text dataset is composed of semantically independent passages, each paired with a query generated to reflect its topic. We use Hit@K as the task-centric metric.

\etitle{\textbf{Hit@K}.} This metric measures whether the most semantic relevant paragraph is included in the top-K retrieved results. Such a self-retrieval task is commonly used in downstream text applications, such as duplicate content detection.
\begin{equation}
\text{Hit}@K = \frac{1}{|\mathcal{Q}|} \sum_{\mathbf{q} \in \mathcal{Q}} \mathcal{I} \left( L(\mathbf{q}) \in \left\{ L(\mathbf{r}) \mid \mathbf{r} \in R(\mathbf{q}) \right\} \right)
\end{equation}
where $\mathcal{Q}$ is queries set, $L(x)$ is a label function to get label for vector $x$, $\mathcal{I}(\cdot)$ is an identity function returning 1 if input statement holds and 0 otherwise, $R(\mathbf{q})$ denotes the top-$K$ retrieved vectors for $\mathbf{q}$.

\eetitle{Embedding Generation.} Since the transformer-based, pretrain \& finetune paradigm has dominated this text embedding training literature, we employed Stella~\cite{stella_en_1_5B} as our text embedding model. This model is tailored for complex text semantic representation and is distilled from a larger, high-performing architecture (i.e., Qwen2~\cite{bai2025qwen2}). It generates 1024-dimension vectors and exhibits strong representational capacity on standard text embedding benchmarks~\cite{muennighoff2023mteb}.

\begin{table*}[tb!]

 \centering
 \caption{Summary of benchmark \vess methods. Build and search complexity are from original papers when available; otherwise, they are empirically estimated. Similar search complexity methods can perform differently in practice due to the absence of proven theoretical bounds (see \S\ref{sec:exp}). Cross Metric reflects the method's ability to seamlessly support multiple similarity metrics within a single index. }
 \label{tab:index_taxonomy}  
 % \small
 \resizebox{0.99\linewidth}{!}{
\begin{tabular}{ccccccc}
\toprule
 \textbf{Similarity Metric} & \textbf{Method} & \textbf{Category} & \textbf{Build Complexity} & \textbf{Search Complexity} & \textbf{Index Storage} & \textbf{Cross Metric}\\
 \midrule
\multirow{7}{*}{\makecell[c]{Inner Product}} 
 & \textsf{Fargo} & Partition-based & $O(nd)$ & $O(\beta nd)$, $\beta \ll 1$ & Low. & \xmark \\
 & \textsf{ScaNN} & Partition-based & $O(knd)$ & $O(c_1\frac{n}{k}d_{quan})$& High. & \xmark \\
 & \textsf{ip-NSW} & Graph-based & $O(nd\log n)$ & $O(d\log n)$ & High. & \xmark \\
 & \textsf{ip-NSW+} & Graph-based & $O(nd\log n)$ & $O(d\log n)$ & High. & \xmark \\
 & \textsf{Mobius} & Graph-based & $O(nd\log n)$ & $O(d\log n)$ & High. & \xmark \\
 & \textsf{NAPG} & Graph-based & $O(nd\log n)$ & $O(d\log n)$ & High.& \xmark \\
 & \textsf{MAG} & Graph-based & $O(nd\log n + dn^{1.16})$ & $O(d\log n)$ & High.& \cmark\\
 \midrule
 \multirow{6}{*}{\makecell[c]{Euclidean Distance}}
 & \textsf{RaBitQ} & Partition-based & $O(2(nd^2+nd))$  & $O(c_1d\log \log d)$ & High. & \xmark \\
 & \textsf{IVFPQ} & Partition-based & $O(knd)$ & $O(c_1\frac{n}{k}d_{quan})$ & High. & \xmark \\
 & \textsf{DB-LSH} & Partition-based & $O(nd)$ & $O(n^{p^*}d \log  n)$, $p^* \ll 1$ & Low. & \xmark \\
 & \textsf{HNSW} & Graph-based & $O(nd\log n)$ & $O(d\log n)$ & High. & \xmark \\
 & \textsf{NSG} & Graph-based & $O(nd\log n + dn^{1.16})$ & $O(d\log n)$ & High. & \xmark \\
 & \textsf{Vamana} & Graph-based & $O(dn^{1.16})$ & $O(dn^{0.75})$ & High. & \xmark \\
\bottomrule
 \end{tabular}}
   \begin{tablenotes}
    \item[a] \small $n$: number of data points; $d$: dimension; $k$: number of clusters; $d_{quan}$: dimension of quantized vectors; $\beta$ and $p^*$: LSH parameter from the origin papers.
  \end{tablenotes}
\vspace{-2mm}
\end{table*}

\stitle{(4) Recommendation} Recommendation is a key application area for \vess. In this setting, \vess is used to retrieve items—such as products, content, or services—that are both semantically relevant to a user's preferences and aligned with broader popularity trends. A fundamental goal of retrieval in recommendation systems is to balance personal relevance with item popularity.

\eetitle{Data Collection.} The \textsf{Commerce} dataset, derived from anonymized traffic logs of a major e-commerce platform, serves as a representative benchmark for large-scale E-commerce systems. Collected over several months, the dataset comprises 99,085,171 records of frequently purchased grocery items. In addition, a query set of 64,111 real user requests was constructed to summarize user profiles and associated intent keywords. Each query is annotated with a set of items with which the users show high preferences, enabling evaluation on downstream recommendation tasks. 

\eetitle{End Task Design.} We use the hit of user preferred Item IDs as labels for \textsf{Commerce}. In this task, \vess is required to retrieve vectors that are both relevant and popular. Thus, we use Matching Score@K as a task-centric metric.

\etitle{\textbf{Matching Score@K}.} It is used to evaluate downstream utility on the Commerce dataset. Because there are usually multi hits for each user, we use the cumulative score of those hits. This metric aligns with practical e-commerce objectives by promoting accurate and relevant product recommendations.
\begin{equation}
    \text{Matching Score}@K = \sum_{\mathbf{q} \in \mathcal{Q}} \sum_{i \in R(\mathbf{q})} \mathcal{I}(L(i) \in \mathcal{H}(\mathbf{q})) \cdot P(L(i))
\end{equation}
Where $\mathcal{Q}$ is the queries set, $L(x)$ is a label function to get the label for vector $x$, $\mathcal{I}(\cdot)$ is an identity function returning 1 if the input statement holds and 0 otherwise, $R(\mathbf{q})$ denotes the top-$K$ retrieved vectors for $\mathbf{q}$, $\mathcal{H}(\mathbf{q}))$ is the annotated items set related to query $q$ and $P()$ is the popular score related to the item.

\eetitle{Embedding Generation.} We used \textsf{ResFlow}, a pre-trained personalized deep retrieval model~\cite{resflow_github}, to generate 48-dimensional embeddings, a size chosen to balance performance and resource limitations in E-commerce platform. Unlike many high-dimensional datasets with low intrinsic dimensionality~\cite{costa2005estimating,lowe2004distinctive}, \textsf{Commerce} exhibits an intrinsic dimension nearly equal to its actual dimension, rendering a high information compression rate.

\subsection{Methods Selection for Evaluation}
\label{sec:algorithm_evaluation}
We focus on two core directions in vector similarity search: Approximate Nearest Neighbor Search (\anns) and Maximum Inner Product Search (\mips), evaluating cutting-edge methods in each. Representative state-of-the-art methods were selected and categorized into partition-based and graph-based methods based on their retrieval mechanisms (Table~\ref{tab:index_taxonomy}).

\stitle{Partition-based Methods.} Partition-based methods divide the vector space based on similarity relationships, aiming to group similar vectors into the same partition to reduce the search space. These methods include tree-based, hash-based, and clustering-based approaches. We excluded tree-based methods from our evaluation due to their performance decline in high-dimensional spaces~\cite{aumuller2020ann}. In \textsf{Iceberg}, we include 5 representative partition-based methods widely used in the fields of \mips and \nns: \textsf{Fargo}~\cite{zhao2023fargo}, \textsf{ScaNN}~\cite{guo2020accelerating}, \textsf{DB-LSH}~\cite{tian2023db}, \textsf{RaBitQ}~\cite{gao2024rabitq}, and \textsf{IVFPQ}~\cite{douze2024faiss}.

\stitle{Graph-based Methods.} We refer to \S\ref{sec:background} for the detail information of graph-based methods. In \textsf{Iceberg}, we include 8 representative partition-based methods widely used in the fields of \mips and \nns: \textsf{HNSW}~\cite{malkov2018efficient}, \textsf{NSG}~\cite{fu2019fast}, Vamana~\cite{jayaram2019diskann}, \textsf{Mobius}~\cite{zhou2019mobius}, \textsf{ip-NSW}~\cite{morozov2018non}, \textsf{ip-NSW+}~\cite{liu2020understanding}, \textsf{NAPG}~\cite{tan2021norm}, \textsf{MAG}~\cite{chen2025stitching}.

\subsection{Reproducibility and Maintenance}
\label{sec:reproducibility}

\stitle{Reproducibility.} Reproducibility is a core principle of iceberg. This helps researchers understand and track the information loss funnel. We provide comprehensive resources and clear guidelines to ensure researchers can easily validate and reproduce our results:

\eetitle{Open-Source Codebase.} All the code used for data generation, algorithm execution, and evaluation is available on anonymous GitHub: \textbf{\href{https://github.com/ZJU-DAILY/Iceberg}{https://github.com/ZJU-DAILY/Iceberg}}. It includes scripts for setting up the environment, running experiments, and analyzing results. To promote transparency and reproducibility, we will host all datasets and embedding models used in this paper on Hugging Face. Users can either download the released datasets directly or reproduce them by applying the released embedding models to the raw data. 

\eetitle{Clear Documentation.} Our GitHub repository provides comprehensive documentation, including installation instructions, data schema, usage examples, and algorithm parameter settings.

\stitle{Maintenance.} Maintaining a robust and relevant benchmark for vector similarity search is an ongoing commitment. Our future plan focuses on two directions: (1) regularly updating with state-of-the-art models, tasks, and retrieval methods to reflect evolving real-world applications, and (2) providing containerized environments for reproducible and hassle-free experimentation.

\eetitle{Guidance for extending to other tasks.} At present, \iceberg primarily focuses on four tasks with large, well-curated public datasets. However, our framework is designed to support a wide range of \textsf{VSS}-based tasks, such as anomaly detection~\cite{li2024nlp} and knowledge graph link prediction~\cite{safavi2020codex}. For new tasks, integration into the framework can be achieved by following the pipeline outlined below: (1) Select an embedding model to encode the raw dataset; (2) Define downstream evaluation metrics for the target task; (3) Apply the \textsf{VSS} method for indexing and retrieval; (4) Conduct evaluation and analysis. We provide additional examples in our GitHub repository.

\section{Experimental Evaluation}
\label{sec:exp}

In this section, we evaluate the end-to-end performance of vector similarity search across a variety of well-structured datasets from multiple domains. Our investigation explores the information loss funnel and the practical efficacy of these systems by addressing four pivotal research questions:
\begin{itemize}[leftmargin=*]

    \item \textbf{RQ-1:} How does information loss funnel impact task-centric metrics utility in vector similarity search applications?
    \item \textbf{RQ-2:} How to select the most suitable vector similarity search algorithm considering the information loss funnel?
    \item \textbf{RQ-3:} What is the realistic ranking of the methods in terms of synthetic metrics and task-centric metrics utility comparison?
    \item \textbf{RQ-4:} What insights can guide the performance optimization of vector similarity search across different scenarios?
\end{itemize}

\subsection{Experiment Setup}

\stitle{Configuration.} The server is equipped with 2 Intel Xeon E5-2650 v4 CPUs running at 2.20 GHz, each with 12 physical cores and hyper-threading enabled, totaling 48 hardware threads. It has 128 GB memory. The operating system is CentOS Linux 7 (Core). We used the official implementations of the evaluated methods, making only necessary modifications for fair testing. We used \textsf{Faiss} 1.10 and \textsf{ScaNN} 1.3.5 for Python calls to \textsf{IVFPQ} and \textsf{ScaNN}, respectively. The other methods were implemented in C++.

\stitle{Synthetic Evaluation Metric}. In \iceberg, we evaluate method performance using both task-centric metrics and synthetic metrics. The task-centric metrics for each task are defined in \S\ref{sec:dataset}. For synthetic metrics, we adopt evaluation protocols commonly used in the \vess literature. Let $R_t'$ denote the set of $K$ vectors returned by the \vess method, and $R_t$ represent the ground truth based on the similarity metric, the synthetic recall@K is formally defined as 
\begin{equation}
  \text{Synthetic Recall} @K = \frac{|R_t \cap R_t'|}{|R_t|} = \frac{|R_t \cap R_t'|}{K}
\end{equation}
synthetic recall is used to evaluate the recall of \vess under a given distance metric, allowing us to assess the trade-off between recall and query speed based on ground-truth similarity.

\begin{figure}[tb!]
\vspace{1ex}
\centering
\centerline{\includegraphics[width=0.9\linewidth]{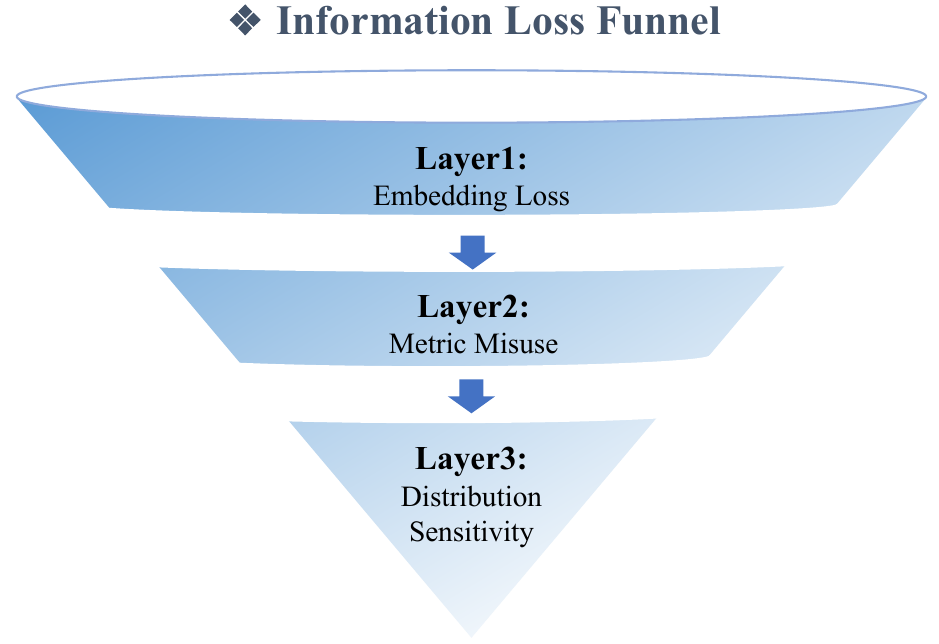}}
\vspace{-2ex}
\caption{Information loss funnel pipeline.}
\vspace{-2ex}
\label{fig:info_funnel}
\end{figure}

\subsection{Impact of the Information Loss Funnel} 
\label{sec:info_loss}

The existence of an information loss funnel is a key reason why improvements in synthetic metrics do not always translate into gains in task-centric metric effectiveness (Figure~\ref{fig:info_funnel}). To better understand this phenomenon, \iceberg conducts comprehensive experiments to systematically investigate three critical sources of information loss in the current end-to-end \vess pipeline. Through this analysis, we aim to address \textbf{RQ1}: {\em How does information loss impact task-centric metrics  utility in vector similarity search applications?} To illustrate our key findings, we present a partial subset of representative results in this section. The full set of experimental results, covering all observation dimensions, is available on our benchmark website.

\begin{table}[tb!]
 \centering
 \caption{Label Recall@100 on selected benchmark datasets. The corresponding synthetic recall@100 is set to 99\%. IP denotes Inner Products and ED is Euclidean distance.}
 \label{tab:lr_performance}
 \resizebox{1\linewidth}{!}{%
 \begin{tabular}{llcc} % Column definition changed to 4 columns
  \toprule
  % Header simplified to a single row
  \textbf{Metric} & \textbf{Method} & \makecell[c]{\textbf{ImageNet-DINOv2} \\ (Label Recall@100)} & \makecell[c]{\textbf{Glint360K-ViT} \\ (Label Recall@100)} \\
  \midrule
  \multirow{7}{*}{\makecell[c]{\textsf{IP}}}
  & \textsf{Fargo}     & 71.38\% & 90.3\% \\
  & \textsf{ScaNN}     & 71.42\% & 90.5\% \\
  & \textsf{ip-NSW}    & 71.4\%  & 90.4\% \\
  & \textsf{ip-NSW+}   & 71\%    & 90.3\% \\
  & \textsf{Mobius}    & 71.3\%  & 88.9\%     \\
  & \textsf{NAPG}      & 71.4\%  & 90.38\%     \\
  & \textsf{MAG}       & 71.4\%  & 90.5\%     \\
  \midrule
  \multirow{6}{*}{\makecell[c]{\textsf{ED}}}
  & \textsf{RaBitQ}    & 71.2\%      & 90.3\%     \\
  & \textsf{IVFPQ}     & 71.2\%      & 89\%     \\
  & \textsf{DB-LSH}    & 71\%      & 90.2\%     \\
  & \textsf{HNSW}      & 71.17\%      & 90.29\%     \\
  & \textsf{NSG}       & 71.17\%      & 90.29\%     \\
  & \textsf{Vamana}    & 71\%      & 89.3\%     \\
  \bottomrule
 \end{tabular}}
 \vspace{-2mm}
\end{table}

\begin{figure}[tb!]
\vspace{1ex}
\centering
\centerline{\includegraphics[width=1\linewidth]{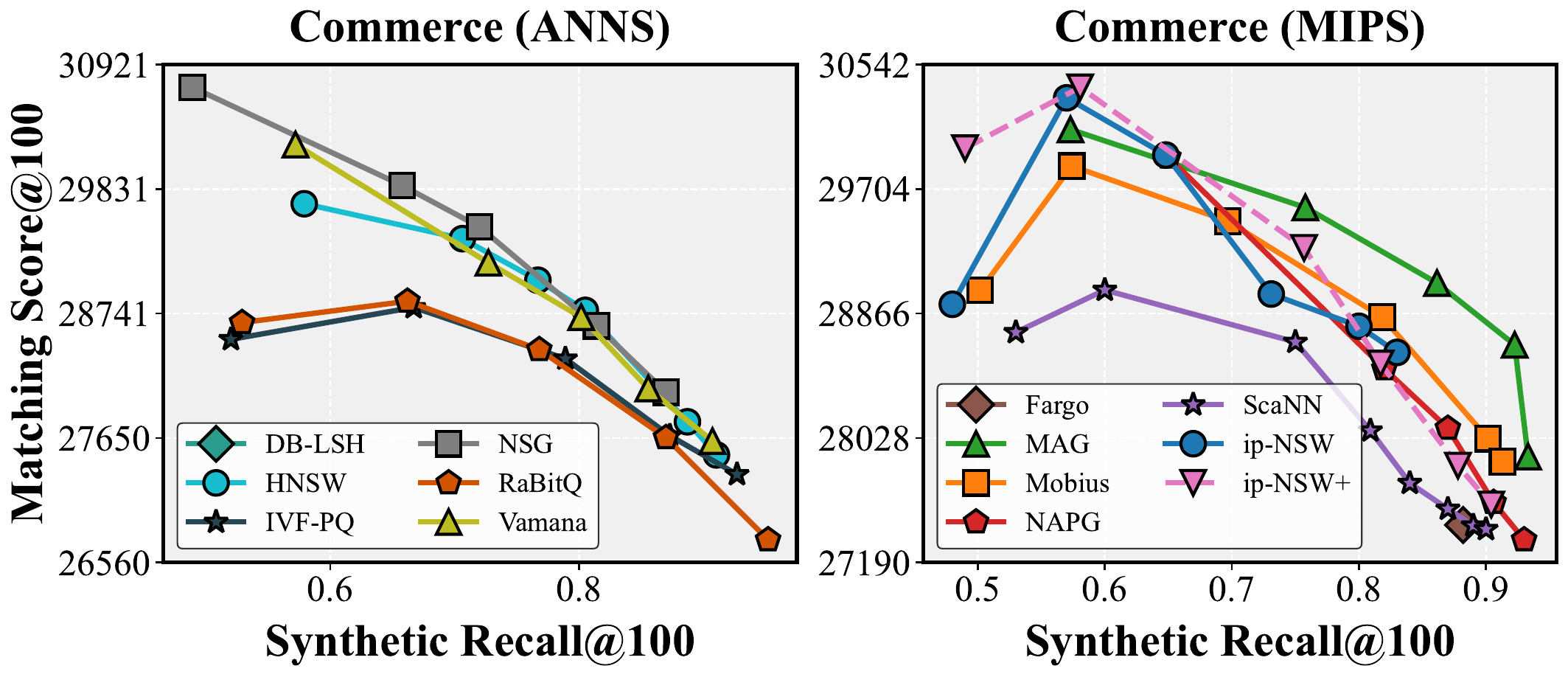}}
\vspace{-2ex}
\caption{Comparison between synthetic and task-centric matching score on \textsf{Commerce} Dataset.}
\vspace{-2ex}
\label{fig:o2.2}
\end{figure}

\stitle{O1 (Embedding Process):} \textbf{Information loss first occurs during the transformation from raw data to embeddings. This loss can untie the semantic similarity with spatial closeness under given distance metrics, making the pursue for high synthetic recall meaningless.} Our experiments reveal a consistent gap between synthetic metrics and true downstream task performance. \textbf{First}, as shown in Table~\ref{tab:lr_performance}, we observe that downstream utility (label recall@100) exhibits a natural upper bound, even when synthetic recall approaches 100\%. For instance, on the \textsf{ImageNet} dataset with \textsf{DINOv2} embeddings, label recall@100 saturates at 71\% despite vector recall@100 reaching 99.9\%. A similar bottleneck is observed in the \textsf{Glint360K-ViT} dataset, where label recall@100 plateaus at 91\% while synthetic recall is nearly perfect. This indicates that high \vess synthetic accuracy does not necessarily translate into better task performance. \textbf{Second}, on the \textsf{Commerce} dataset, we find a non-monotonic relationship between synthetic metrics accuracy and downstream task-centric utility (Figure~\ref{fig:o2.2}). As synthetic recall improves, the matching score@100—a task-specific measure of user interest—initially increases but then declines. This counterintuitive trend shows that relying too heavily on synthetic measures without considering their practical implications can lead to misleading conclusions. \textbf{Finally}, Table~\ref{tab:o2.3} shows that downstream task-centric utility is highly sensitive to the choice of embedding model. Using \textsf{DINOv2}, \textsf{EVA02}, \textsf{ConvNeXt}, and \textsf{AlexNet} on ImageNet, we observe significant differences in label recall@100 under the same retrieval setting (all with 99\% synthetic recall). While \textsf{DINOv2} achieves only 71\%, \textsf{EVA02} and \textsf{ConvNeXt} reach 85\% and 84\%, respectively. Furthermore, representations from the old embedding model diverge more from task-centric retrieval targets. \textsf{AlexNet}-based embeddings achieved only about 21\% accuracy on label recall, suggesting that neglecting downstream evaluation on outdated embeddings leads to increasingly misaligned and less meaningful retrieval results.

We attribute this phenomenon to information loss in the representation space, which is closely related to concepts such as \textbf{neural collapse}~\cite{papyan2020prevalence} and \textbf{representation collapse}~\cite{chi2022representation,li2022understanding}. Such representation collapse reflects an Occam's razor principle: the model will only retain information useful for predicting the labels. However, in many domains of machine learning, the optimization objective of the model doesn't always facilitate a metric-preserving projection over the whole sample space. For example, classification loss only fits the training data to the training labels, suffering from generalization errors to test space. Contrastive learning aligns the metric structure of extracted vectors among selected positive and negative samples (usually applying in-batch sampling), not the whole sample space. After training on a dataset, different amount of information will be lost according to different loss design, needless to say the model capacity. To focus on \vess, we don't aim to propose solutions to this problem and leave it to the machine learning community.

\begin{figure*}[tb!]
\vspace{1ex}
\centering
\centerline{\includegraphics[width=1\linewidth]{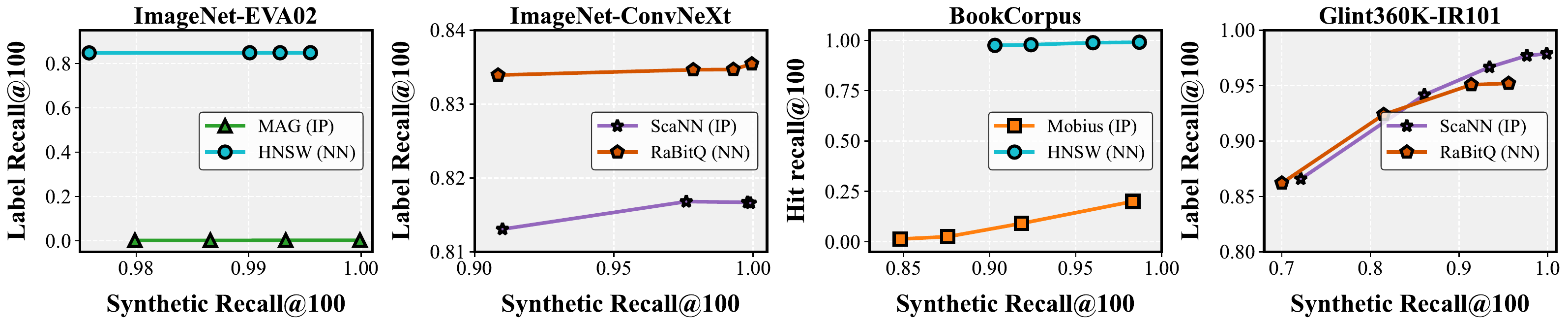}}
\vspace{-2ex}
\caption{Comparison of downstream task-centric performance under different similarity metrics.}
\vspace{-2ex}
\label{fig:o3.1}
\end{figure*}

\begin{figure*}[tb!]
\vspace{1ex}
\centering
\centerline{\includegraphics[width=1\linewidth]{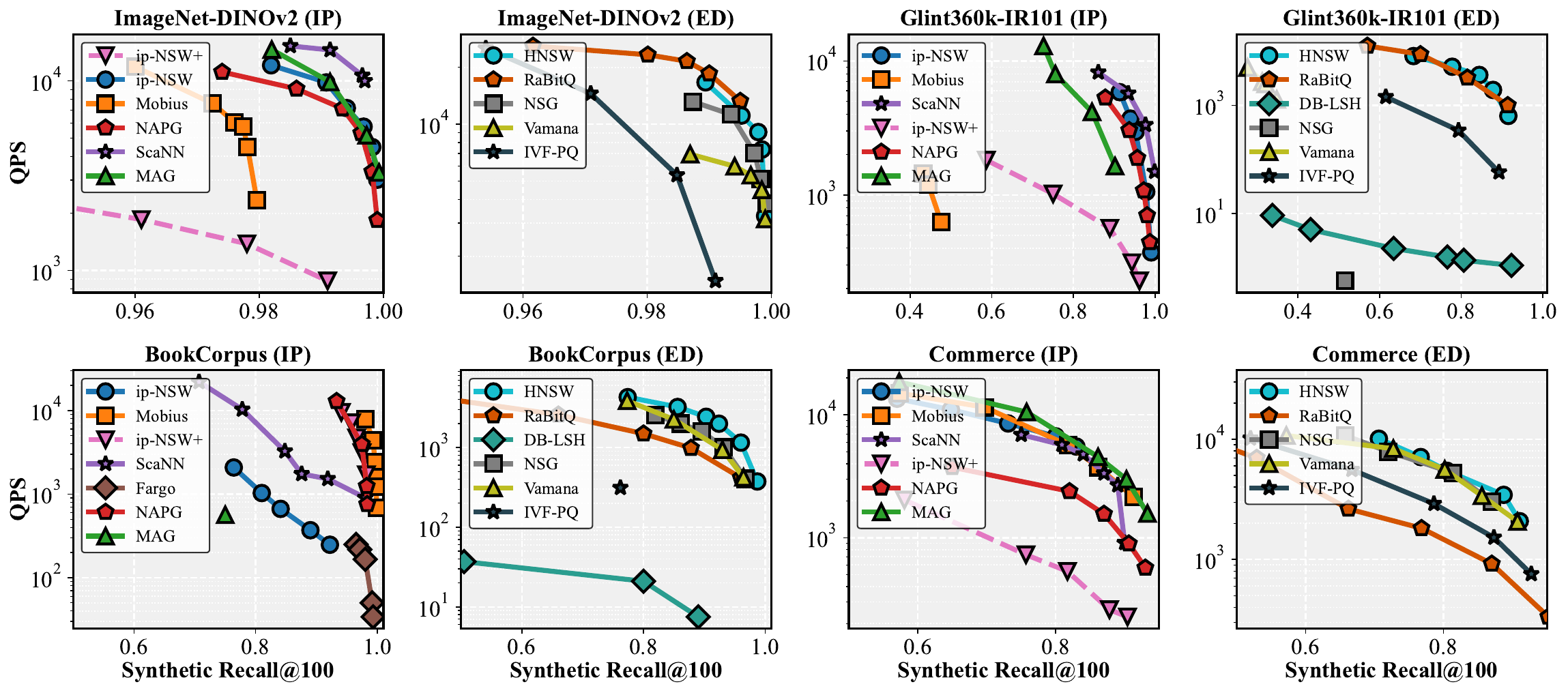}}
\vspace{-2ex}
\caption{Experimental results of query performance for synthetic metric recall on four representative datasets.}
\vspace{-2ex}
\label{fig:o4.1}
\end{figure*}

\stitle{O2 (Metric Choice):} \textbf{The second layer of the information loss funnel is metric misuse, a critical issue in real-world applications where inappropriate similarity metrics for \vess render even high synthetic recall ineffective for task-centric gains.}  This issue is prevalent and often overlooked when evaluating \vess systems solely based on synthetic metrics. Figure~\ref{fig:o3.1} shows the trend of task-centric performance as synthetic recall varies on representative datasets, \textbf{using the best-performing methods from both the \anns and \mips domains}.

Figure~\ref{fig:o3.1} reports results across four datasets: \textsf{ImageNet-EVA02}, \textsf{ImgeNet-ConvNeXt}, \textsf{BookCorpus}, and  \textsf{Glint360-IR101}, demonstrating how the similarity metric choice radically impacts downstream task-centric performance.
On \textsf{ImageNet}, with \textsf{ConvNeXt} embeddings, \anns using Euclidean distance scores roughly 4\% higher label recall@100 than \mips, even though both are run at synthetic recall@100 around 99.9\%. The problem is more pronounced with the \textsf{EVA02} model: even though \mips achieves nearly perfect synthetic recall@100 (99.99\%), actual task-centric label recall@100 is stubbornly below 1\%, reflecting a near-complete semantic mismatch. Analogously, on the text retrieval dataset \textsf{BookCorpus}, \mips manages less than 50\% task-centric hit recall, whereas Euclidean-based \anns attains 100\%. However, the \textsf{Glint360-IR101} dataset shows the opposite. Here, \mips comfortably outperforms Euclidean-based \anns on task-centric metrics: with \textsf{Glint360K-IR101} embeddings, \mips performs 98\% label recall@100, whereas \textsf{HNSW} (\anns methods) tops out at 92\%. Given the high-stakes nature of applications such as identity verification, such performance discrepancies result in critical system failures—even with high synthetic recall.

These differences in performance are due to mismatched similarity measures and embedding space geometries, controlled through \textbf{loss design} and \textbf{metric affinity}. Models such as \textsf{ConvNeXt} are trained with cross-entropy loss, which promotes embeddings within the same class to have tight clusters in Euclidean space. Euclidean distance naturally captures this structure. By contrast, \mips is sensitive to both direction and norm of vectors and does not respect this clustering, retrieving semantically unrelated neighbors despite high synthetic recall. \textsf{Glint360K-IR101} embeddings are trained with \textsf{ArcFace}, which employs an Additive Angular Margin Loss to foster a highly discriminative angular structure—small intra-class angles and large inter-class angles. This structure is naturally captured by inner product similarity, and thus \mips is the correct choice and we see why it has superior task-centric performance on \textsf{Glint360}. Combined, these results highlight an important takeaway: Evaluating synthetic recall with a single similarity metric (like Euclidean distance) can be misleading and, unless that metric is aligned with the training loss, largely meaningless. Metric choice materially affects outcomes: metric misuse not only incurs drastic information loss but also can make high-synthetic recall \vess systems functionally useless in real-world downstream tasks.

\stitle{O3 (Data Distribution Sensitivity): Each data distribution has methods best suited to its properties or structures, and choosing an unsuitable method can cause retrieval efficiency to drop and result in information loss, contributing the third layer of information loss funnel.} We categorize the methods into two classes—partition-based and graph-based methods (\S\ref{sec:algorithm_evaluation} for details)—and conduct extensive evaluations on four datasets from different domains for both \anns and \mips tasks. The experimental results are presented in Figure~\ref{fig:o4.1}.

\eetitle{O-3.1: Partition- and graph-based methods have their own strengths.} In traditional benchmarks~\cite{wang2021comprehensive, aumuller2020ann}, it is commonly assumed that graph-based methods significantly outperform other types of methods in in-memory settings. This belief has led many commercial vector search systems to prioritize graph-based methods by default (e.g., \textsf{AnalyticDB}~\cite{wei2020analyticdb}), regardless of the application. However, this assumption does not always hold true. A \vess method's efficiency heavily relies on the data distribution. Experimental results demonstrate query performance on a synthetic metric.

As shown in Figures~\ref{fig:o3.1}, the partition-based methods exhibit a clear performance advantage in certain datasets, particularly on the \textsf{ImageNet-DINOv2} and \textsf{Glint360K-IR101}. In the \mips setting, \scann significantly outperforms the leading graph-based method on these datasets, achieving speedups of 1.7$\times$ and 3$\times$, respectively, while maintaining comparable synthetic metric recall. A similar advantage is observed in the \anns setting: on \textsf{Glint360K-IR101}, \textsf{RaBitQ} delivers nearly 2$\times$ faster query speed than \textsf{HNSW} at equivalent synthetic metric recall. In contrast, graph-based methods perform better on \textsf{BookCorpus} and \textsf{Commerce}. In the \mips experiments on synthetic metric, the top-performing graph-based index achieves up to 3$\times$ faster speed than \scann on \textsf{Commerce}, and up to 4$\times$ on \textsf{BookCorpus}. A similar trend is seen in the \anns setting: \textsf{HNSW} outpaces the strongest partition-based competitor, \textsf{RaBitQ}, by 2.7$\times$ and 3.2$\times$ on \textsf{BookCorpus} and \textsf{Commerce}, respectively.
\begin{table}[tb!]
\centering
\caption{Comparison of task-centric performance (label recall@100) across different models on \textsf{ImageNet} dataset.}
\label{tab:o2.3}
\resizebox{\linewidth}{!}{%
\begin{tabular}{@{}lcccccc@{}}
\toprule
\multirow{2}{*}{\textbf{Model}} & \multicolumn{3}{c}{\textbf{HNSW}} & \multicolumn{3}{c}{\textbf{RaBitQ}} \\
\cmidrule(lr){2-4} \cmidrule(lr){5-7}
& \multicolumn{3}{c}{Synthetic Recall@100} & \multicolumn{3}{c}{Synthetic Recall@100} \\
& 90\% & 95\% & 99\% & 90\% & 95\% & 99\% \\
\midrule
\textsf{AlexNet} & 21.62\% & 21.48\% & 21.43\% & 22.55\% & 22.43\% & 22.41\% \\
\textsf{DINOv2} & 70.5\% & 71.08\% & 71.17\% & 70.08\% & 71.2\% & \textbf{71.24\%} \\
\textsf{EVA02} & 80.1\%  & 83.7\% & \textbf{84.9\%} & 84.4\% & 84.76\% & 84.82\% \\
\textsf{ConvNeXt} & 82.5\% & 82.7\%  & 83.3\% & 83.46\% & 83.46\% & \textbf{83.54\%} \\
\bottomrule
\end{tabular}}
\vspace{-2mm}
\end{table}

The performance differences are attributed to the sensitivity of the methods to data properties. For example, when shapes in the data are strongly cluster-like, graph-based methods are liable to get trapped in local optima among clusters, which discourages them from finding the relevant areas~\cite{ruan2025empowering}. Similarly, when angles between vectors are large, edge pruning methods based on the triangle inequality (e.g., \textsf{HNSW}, \textsf{NSG}) increasingly fail to sparsift the graph while maintaining connectivity. Partition-based methods are inherently dependent on data characteristics as well: whether they use clustering or hash-based projections, their quality can significantly vary from one dataset to another~\cite{gaohigh}.

\eetitle{O-3.2: Wrong choice leads to performance degradation.} Beyond the query speed disparities on synthetic metric, the impact of selecting an unsuitable method is more critical—it can constrain synthetic metric recall and downstream task-centric performance, ultimately causing significant information loss.  This phenomenon is particularly pronounced in the domain of \mips. Due to the lack of rigorous theoretical foundations, some methods suffer from synthetic metric recall bottlenecks and performance degradation on certain datasets. For example, while the \textsf{Mobius} algorithm performs well on some datasets (e.g., \textsf{BookCorpus}), it fails to exceed 50\% accuracy on synthetic metric recall@100 on the \textsf{Glint360K-IR101} dataset. Similarly, partition-based methods such as \scann exhibit noticeable performance drops on the \textsf{Commerce} dataset when the target synthetic metric recall exceeds 90\%. These findings highlight that variations in data distribution can lead to different degrees of information loss across methods.

\stitle{Information Loss Funnel Analysis.} Having identified the information loss funnel’s layered structure through extensive analysis, we qualitatively assess each layer’s contribution. \textbf{Layer 1} (Embedding loss) defines the upper bound of task-centric performance, as the transformation from raw data to embeddings inevitably discards fine-grained semantics. \textbf{Layer 2} (Metric misuse) exerts a decisive impact on retrieval accuracy, where metric misalignment undermines the correlation between synthetic recall and task-centric metric. \textbf{Layer 3} (Distribution sensitivity) primarily shapes retrieval efficiency, though its impact on task-centric metrics, albeit smaller, remains non-trivial. Overall, the first two layers dominate task-centric accuracy, while the third governs computational scalability.
 We now revisit key \vess applications' information loss within this evaluation framework. 

\eetitle{Image Classification.} \textbf{Layer 1:} Image classification models, both supervised and self-supervised, extract representations but do not achieve perfect task-centric accuracy due to data complexity and model limitations (Table~\ref{tab:o2.3}). \textbf{Layer 2:} Representations in image classification tasks tend to align well with Euclidean space, as their embedding models are typically trained with cross-entropy loss, which encourages embeddings of the same class to form compact clusters in Euclidean space. \textbf{Layer 3:} Different methods show inconsistent performance between synthetic and downstream tasks, leading to information loss in method selection.

\eetitle{Face Recognition.} \textbf{Layer 1:} ArcFace-based models may encounter a task-centric accuracy ceiling in downstream tasks due to architectural constraints on representation learning (Table~\ref{tab:lr_performance}). \textbf{Layer 2:} Inner product metrics are better suited for face recognition due to the use of Additive Angular Margin Loss, which promotes a discriminative angular structure. \textbf{Layer 3:} Graph-based methods are more prone to local optima than partition-based methods, reducing their effectiveness on synthetic metric tasks (Figure~\ref{fig:o4.1}).

\eetitle{Text Retrieval.} \textbf{Layer 1:} Current text representation models exhibit some distortion when extracting information from paragraphs. \textbf{Layer 2:} Incorrect metric choice, such as inner product, can reduce task-centric Hit@K performance to under 50\% (Figure~\ref{fig:o3.1}). \textbf{Layer 3:} Due to the difficulty in distinguishing boundary points, partition-based methods are more prone to information loss in synthetic task retrieval than graph-based methods.

\eetitle{Recommendation.} \textbf{Layer 1:} \textsf{ResFlow} struggles to strongly correlate synthetic and task-centric tasks, leading to significant information loss (Figure~\ref{fig:o2.2}). \textbf{Layer 2:} Euclidean distance and inner product metrics show an initial performance improvement followed by a decline in downstream tasks, because models are trained on historical user–item interactions, a user vector's nearest items are typically those already purchased; with repurchases rare, optimizing for such neighbors yields limited real-world benefit. \textbf{Layer 3:} Partition-based methods suffer retrieval performance degradation in synthetic metrics (Figure~\ref{fig:o4.1}). 

\subsection{Decision Tree For Method Selection}

\label{sec:data_align}
According to above funnel analysis, we still wish to propose a guide in \vess method selection even with hands tied (embedding model capacity bottleneck). Here we delve into the philosophy behind selecting an appropriate similarity metric and \vess method under the framework of the information loss funnel, aiming to addressing \textbf{RQ2}: {\em How to select the most suitable vector similarity search method considering the information loss funnel?} There are two key factors in choosing the best method: one is selecting the metric that best aligns with the downstream task, and the other is choosing the method that performs best under that metric. To operationalize this, we construct a two-layer interpretable decision tree model based on four meta-features. In the first layer, we prioritize metric selection, as aligning the metric with the downstream task is critical for preserving task-centric accuracy. In the second layer, we use data distribution-related meta-features to further distinguish between methods, enabling the selection of the most effective method for each dataset.  The meta-features are in Table~\ref{tab:metric}.

\stitle{O4 (First Layer: Selecting Metric):} \textbf{The initial layer of our decision tree focuses on selecting the most suitable similarity metric for a given embedding space: either \ip or Euclidean distance.} This choice is critical as it directly impacts the downstream task-centric utility. Given the discussion in \S\ref{sec:info_loss}, the embedding model domain evolves fast and usually does not offer clues on how is the affinity of their method with downstream retrieval tasks. However, with exhaustive research, we manage to capture affinity patterns between data distribution property or structure with different metrics used for \vess. Our decision rule for metric selection is based on the combined assessment of the \textsf{DBI} and \textsf{CV} as follows.

\begin{itemize}[leftmargin=*]
   \item \textbf{Davies-Bouldin Index} (\textsf{DBI}): \textsf{DBI} measures cluster compactness and separation in the embedding space; lower values indicate better clustering quality and can be computed using either Euclidean distance (\textsf{DBI}$_E$) or cosine similarity (\textsf{DBI}$_C$).
    \begin{equation}
    \textsf{DBI} =\frac{1}{N}\sum_{i=1}^N\max_{j\neq i}\frac{\sigma_i+\sigma_j}{d(c_i,c_j),}
    \end{equation}
    where $N$ denotes the number of clusters, $\sigma_i$ denotes the intra-cluster distance of cluster $i$, $c_i$ and $c_j$ denote cluster centers, and $d(c_i,c_j)$ denotes the distance between cluster centers.

    \item \textbf{Coefficient of Variation }(\textsf{CV}): It quantifies embedding vector norm variability; a lower CV indicates more consistent norms. This is typical in training with norm normalization (e.g., contrastive learning), which mitigates the impact of vector norms on model behavior.
    \begin{equation}
      \text{CV} = \sigma(\lVert x\rVert)/\mathbb{E}(\lVert x\rVert).   
    \end{equation}
    Here $\sigma(\lVert x\rVert)$ is the standard deviation of vector norms. $\mathbb{E}(\lVert x\rVert)$ is the mean norm over the dataset. 
\end{itemize}
\begin{table}[tb!]
  \caption{The detailed value of four meta-features for benchmark datasets.}
  \label{tab:metric}
  \resizebox{0.99\linewidth}{!}{
  \begin{tabular}{c|c|c|c|c|c}
    \toprule
    \textbf{Dataset} & \textbf{DBI$_E$} & \textbf{DBI$_C$} & \textbf{CV}  & \textbf{RA} & \textbf{RC} \\
    \midrule
    \textsf{ImageNet-DINOv2} &  3.04 & 1.61 & 0.02 & 83  & 1.98 \\
    \textsf{ImageNet-EVA02} & 5.35 & 6.42 & 0.02 & 3 & 1.59\\
    \textsf{ImageNet-ConvNeXt} & 1.4 & 1 & 0.36 & 81 & 3.6 \\
    \textsf{Glint360K-ViT} & 4.27 & 2.26 & 0.04 & 91 & 142 \\
    \textsf{Glint360K-IR101} & 5.51 & 2.09 & 0.08 & 87 & 1.38\\
    \textsf{BookCorpus} & 4.5 & 9.7 & 0.13  & 44 & 75\\
    \textsf{Commerce} & 2.68 & 2.35 & 0.00 & 38 & 3\\
    \bottomrule
  \end{tabular}}
\end{table}
\begin{table*}[tb!]
    \centering
    \caption{New Method Leaderboard on \iceberg 1.0. The table is divided into synthetic recall ranking and task-centric ranking evaluations. For the synthetic recall Ranking, we compare the query speed required to achieve the same level of synthetic recall. In contrast, for the task-centric ranking, we compare the query speed required to achieve the same level of task-centric metric.}
    \label{tab:ranking}
    \vspace{-1mm}
    \resizebox{\linewidth}{!}{
    \begin{tabular}{l cl cl | ccl}
        \toprule
        % Top-level header now spans 4 columns for the first group
        & \multicolumn{4}{c|}{\textbf{Synthetic Recall Ranking}} & \multicolumn{3}{c}{\textbf{Task-Centric Ranking}} \\
        \cmidrule(r){2-5} \cmidrule(l){6-8}
        & \multicolumn{2}{c}{\textbf{ANNS (for ED)}} & \multicolumn{2}{c|}{\textbf{MIPS (for IP)}} & & & \\
        \cmidrule(r){2-3} \cmidrule(r){4-5}
        \textbf{Datasets} & \textsc{method Type} & \textsc{Winner} & \textsc{method  Type} & \textsc{Winner} & \textsc{Metric} & \textsc{method Type} & \textsc{Winner} \\
        \midrule
        \textsf{ImageNet-DINOv2}  & \parttype & \textsf{RaBitQ} & \parttype & \scann & \ip & Partition & $\star$ \scann \\
        \textsf{ImageNet-EVA02}   & \graphtype & \hnsw & \graphtype & \magg & \textsf{ED} & Graph    & $\star$ \hnsw \\
        \textsf{ImageNet-ConvNeXt} & \graphtype & \textsf{HNSW} & \parttype & \scann & \textsf{ED} & Partition & $\star$ \textsf{RaBitQ} \\
        \textsf{Glint360K-ViT}    & \graphtype & \hnsw & \parttype &  \scann & \ip & Partition & $\star$ \scann \\
        \textsf{Glint360K-IR101}  & \parttype & \textsf{RaBitQ}  & \parttype &  \scann & \ip & Partition & $\star$ \scann \\
        \textsf{BookCorpus}       & \graphtype &  \hnsw & \graphtype & \textsf{Mobius} & \textsf{ED} & Graph  & $\star$ \hnsw \\
        \textsf{Commerce}   & \graphtype & \textsf{VAMANA} & \graphtype &  \magg  & \ip & Graph     & $\star$ \textsf{ip-NSW+} \\
        \bottomrule
    \end{tabular}
    }
\end{table*}

\textbf{The datasets opt for \ip as the similarity metric when the condition $(DBI_E \ge DBI_C) \wedge CV \leq 0.1$ is satisfied.} This condition suggests that clusters in the embedding space are more distinguishable in angular terms than in Euclidean space, and that the embedding vectors exhibit relatively uniform norms. This distribution prioritizes angular information, indicating semantic similarity and supporting \mips more effectively, meaningfully retaining task-relevant details downstream. Our findings corroborate this: on the \textsf{Glint360K-IR101} dataset ($DBI_E = 5.51$, $DBI_C = 2.09$, $CV = 0.08$), \ip achieves a 4\% higher label recall@100 compared to \anns, which is a notable boost for face recognition. Otherwise, we opt for Euclidean distance instead. In that case, the data has more Euclidean clustering, and high \textsf{CV} indicates there is large norm variation in vectors. This biases $\ip$ similarity towards norm variations instead of semantic relevance and downstream task performance thus suffers.

\stitle{O5 (Second Layer: Selecting Algrithm):} \textbf{Our second layer thus chooses the best \vess method for an embedding space and a chosen similarity metric.} Mainstream \vess engines predominantly employ either partition-based or graph-based methods, which differ significantly in their strategies. However, method choice is connected to information loss and data distribution structures too. Our decision rule is subject to \textsf{RA} and \textsf{RC}.

\begin{itemize}[leftmargin=*]

    \item \textbf{Relative Angle} (\textsf{RA}): It measures the average angular relationships between embedding vectors, where a larger \textsf{RA} value indicates greater angular dispersion within the dataset. 
    \begin{equation}
        \text{RA} = \frac{1}{N} \sum_{i=1}^{N} \arccos\left( \frac{\mathbf{v}_i \cdot \mathbf{C}}{\|\mathbf{v}_i\| \|\mathbf{C}\| + \epsilon} \right)
    \end{equation}
    where $\mathbf{v}_i$ is an embedding vector from the dataset of $N$ vectors, $\mathbf{C} = \frac{1}{N} \sum_{i=1}^{N} \mathbf{v}_i$ is the global mean vector, and $\epsilon$ is a small positive constant for numerical stability.
    \item \textbf{Relative Contrast} (\textsf{RC}): It measures a dataset's relative density in high-dimensional space; higher values suggest a more dispersed distribution.
    \begin{equation}
        \text{RC} = \frac{1}{N} \sum_{i=1}^{N} \frac{D_{\text{mean}}^{v_i}}{D_{\min}^{v_i}}
    \end{equation}
    Here, $D_{\min}^{v_i}$ represents the distance from $\mathbf{v_i}$ to its approximate nearest neighbor, and $D_{\text{mean}}^{v_i}$ is the mean distance from $\mathbf{v_i}$ to vectors (random sampled) in dataset.
\end{itemize}

\textbf{The datasets favor partition-based methods when the condition $\text{RA} \ge 60 \vee \text{RC} \leq 1.5$ is met}. Both \textsf{RA} and \textsf{RC} play critical roles in determining the search navigability of graph-based indexing methods. A low \textsf{RA} indicates strong cluster segregation, leading to poor global connectivity in the graph and causing searches to get stuck in local optima. High angular dispersion further weakens triangle inequality pruning, degrading proximity graph quality. Similarly, a low \textsf{RC} suggests that local neighbors have similar distances, reducing structural discriminability and diminishing the effectiveness of graph-based guidance during search. In such conditions, partition-based methods perform more efficiently. By detecting spatial divisions in the data, they restrict computation to relevant regions during querying and benefit from batch processing. As shown in Figure~\ref{fig:o3.1}, on the \textsf{Face-IR101} dataset (\textsf{RA} = 87, \textsf{RC} = 1.38), \scann achieves 3$\times$ lower query latency than the state-of-the-art graph-based method \textsf{ip-NSW}, while maintaining comparable synthetic recall.
Conversely, graph-based methods excel when \textsf{RA} is high and \textsf{RC} is moderate. Better connectivity and sparsity enable effective edge pruning, and the power-law structure of the graph allows greedy search to traverse efficiently. A high \textsf{RC} improves search quality by reducing redundancy and enhancing guidance. For instance, \hnsw outperforms \textsf{RaBitQ} by 3.7$\times$ in query speed on the \textsf{BookCorpus} at synthetic recall@100 = 0.95.

\stitle{Threshold analysis.} Our distribution-related threshold conclusions are supported by diverse datasets and reported under the confidence interval lower bound, which provides strong tolerance and generalizability. For example, $(\mathrm{CV}_{\text{IP}} = [0.02, 0.04, 0.08, 0.00])$ yielded a 95\% upper confidence bound of \textbf{0.1019}, 
whereas the \textsf{ED} subset $(\mathrm{CV}_{\text{ED}} = [0.36, 0.13, 0.02])$ produced a 95\% lower bound of \textbf{--0.1700}. The overlap between their confidence intervals motivated a conservative threshold of $\textsf{CV} = 0.10$, ensuring class separability. In current \textsf{ML} practice, Transformer-based embeddings have largely unified representation models, and with the emergence of more large-scale labeled datasets, we expect both representation methods and thresholds to further converge.

\stitle{Summary.} Following the decision tree partitioning, each dataset and method can be routed to a specific leaf node, where the most appropriate method is selected accordingly. Our information-loss funnel is designed as a general diagnostic model, and the current decision tree
 is fitted to existing models and datasets. The decision tree structure and dataset-method mapping, is shown in Figure~\ref{fig:overview}.

\subsection{Iceberg Leaderboard 1.0}
In this section, we present a task-centric leaderboard produced by \iceberg. Our evaluation spans all eight benchmark datasets, jointly assessing synthetic metrics and downstream utility to reveal the performance of the methods across these datasets. We aim to address \textbf{RQ3}: {\em What is the realistic ranking of the methods in terms of synthetic metrics and task-centric metrics utility comparison?}

\stitle{Parameter Configuration.} Most methods do not allow the user to explicitly specify a quality target. In the \iceberg, we first use a consistent configuration for key shared parameters across methods of the same type. For graph-based methods, we set the two most critical parameters as follows: \textsf{EFC} (queue length during graph construction) is set to 256, and \textsf{M} (maximum number of edges) is set to 32. For partition-based methods that rely on clustering methods (\scann, \textsf{IVFPQ}, and \textsf{RaBitQ}), we set the number of clusters \textsf{C} to $4\sqrt{N}$, where $N$ is the number of vectors. Then, we carefully apply grid search to fine-tune the method-specific parameters based on the recommended settings from the original papers. The detailed settings can be found on the framework’s website.

\stitle{O6 (Leaderboard): } \textbf{\iceberg integrates downstream utility into the evaluation, delivering an end-to-end method ranking that serves as an intuitive and actionable method selection guidance for users. We observe that the realistic performance rankings do not always align with those derived solely from synthetic metrics.} For example, although \textsf{MAG} attains higher synthetic recall at lower latency, its best task-centric score still lags behind \textsf{ip-NSW+}, making the latter the overall winner. The leaderboard is summarized in Table~\ref{tab:ranking}. Users may reproduce these results by selecting the optimal algorithm along the synthetic recall–QPS and task-centric recall–QPS curves.

\subsection{Insights \& Future Directions}
\label{sec:insight}
In this section, we start from the perspective of the information loss funnel and leverage visualization analysis techniques to further explore insights behind the performance improvements in vector similarity search. Our analysis focuses on three key aspects: \textbf{task-aware, metric-aware, and distribution-aware} perspectives, aiming to answer \textbf{RQ4}: {\em What insights can guide the performance optimization of vector similarity search across different scenarios?}

\begin{figure}[tb!]
\vspace{1ex}
\centering
\centerline{\includegraphics[width=1\linewidth]{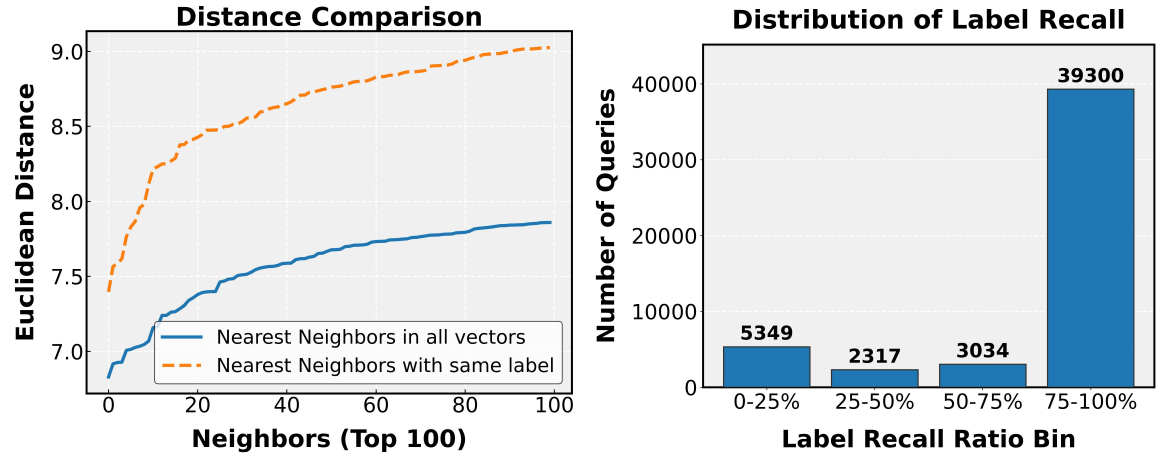}}
\vspace{-2ex}
\caption{Left: Distance distribution between queries and same-label points vs. all database points. Right: Label recall distribution when synthetic metric recall@100=99\%.}
\vspace{-2ex}
\label{fig:o7}
\end{figure}

\stitle{O7 (Task-aware Improvement):} \textbf{Task-aware \vess can help align vector similarity search with downstream tasks.} As shown in Figure~\ref{fig:o7}, the left part illustrates that in the \textsf{ImageNet-ConvNeXt} dataset, the nearest neighbors of a query based solely on Euclidean distance are not necessarily those sharing the same label. In some cases, instances with the same label are located relatively far from the query. The right part presents a global statistical view, revealing that a notable portion of queries fail to exhibit well-aligned distance-label relationships. These observations suggest that further development of task-aware \vess is promising. 

\eetitle{Future Direction.}  Given the diversity of real-world applications, future work should focus on developing adaptable, task-aware \vess techniques that generalize across models and tasks while minimizing system overhead. For example, to address this issue, we can incorporate a \vess method with a task-aware early-stop strategy where label-recall saturates or focus the search within a task-aware proximity range. A more promising future work is to co-optimize indexing and search methods with downstream tasks in mind. When optimizing both the representation learning, \vess components, and downstream tasks together in an end-to-end framework, it is possible to more closely align learned embeddings and \vess mechanisms, make downstream tasks more effective, and minimize information loss accumulated along the pipeline. Such a synergistic design is particularly important in complex, large-scale \textsf{AI} applications.

\stitle{O8 (Metric-aware Improvement):} \textbf{Metric-aware \vess is a breakthrough approach to enhancing \vess performance.} \iceberg has identified the link between similarity metrics and downstream task effectiveness. However, most existing methods are tailored to a single metric, limiting their adaptability. Recent work such as the \magg~\cite{chen2025stitching} demonstrates initial progress by supporting both \ip and \nn within a single index, enabling cross-metric retrieval and improving the usability of \vess. Nevertheless, these methods still rely on manual tuning and cannot fully bridge the geometric differences across metric spaces.

\eetitle{Future Direction.} A key future direction for \vess is developing metric-aware methods that can adapt to various similarity spaces—inner product, Euclidean, and non-Euclidean—without manual tuning. As embedding objectives and architectures grow more diverse and model training does not necessarily indicate which distance metric should be used, fixed-metric approaches become increasingly suboptimal. Overcoming this "metric barrier" calls for self-adaptive methods capable of automatically identifying and adjusting to the appropriate similarity metric. While \magg marks early progress, their reliance on manual tuning and limited support for complex metrics highlights the need for more flexible solutions, especially given their performance gap with metric-specific methods like \textsf{ScaNN} on certain datasets.

\textcolor{blue}{}

\begin{figure}[tb!]
\vspace{1ex}
\centering
\centerline{\includegraphics[width=1\linewidth]{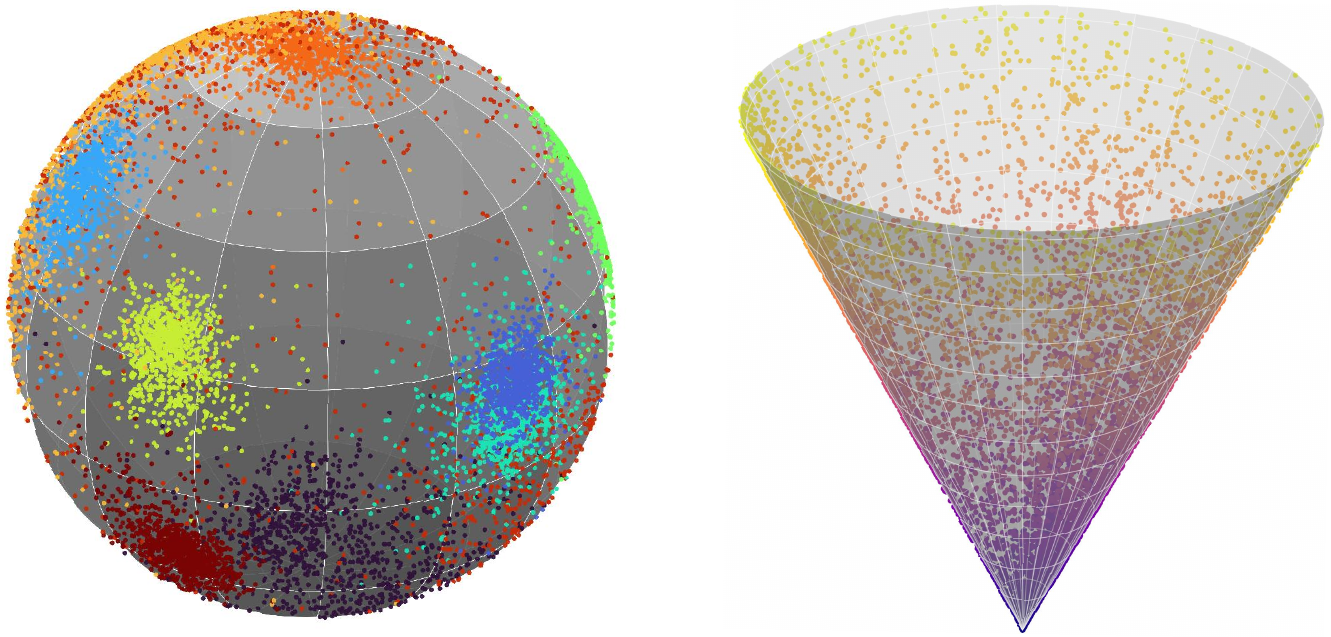}}
\vspace{-2ex}
\caption{Visualization of data distributions: Left – Glint360K-ViT dataset; Right – BookCorpus dataset.}
\vspace{-2ex}
\label{fig:o9}
\end{figure}

\stitle{O9 (Distribution-aware Improvement)} 
\textbf{The heterogeneous pattern distributions of diverse vector datasets inherently challenge traditional \vess methods that rely on fixed construction strategies.} Such methods have poor generalizability to datasets with varied geometric properties. For example, in the \textsf{Glint360K} dataset, as shown in Figure~\ref{fig:o9}(left), K-means clustering is performed in the original space, followed by \textsf{PCA} methods for \textsf{3D} visualization, where data points are separated by cluster labels. The dataset showcases extreme angular separation: inter-class distances are large, while intra-class distances have high variability, thus severely crippling the usefulness of graph-based indices. For the \textsf{BookCorpus} dataset, an analogous dimensionality reduction method was used, with the data points colored according to their norms. Figure~\ref{fig:o9}(right) shows that the average angle between vectors is kept within 60 degrees, and the data is scattered along multiple norm directions. In such scenarios, graph-based methods can effectively leverage the triangle inequality for pruning edges. However, partition-based approaches struggle to create clear boundaries between partitions, making it difficult to exactly identify border points. This limitation negatively impacts their ability to achieve high synthetic metric recall efficiently.

\eetitle{Future Direction.} A key direction for future research lies in designing distribution-aware \vess methods that can adapt to the geometric characteristics of data. Since we can identify effective statistical indicators to divide different methods into different nodes of the decision tree (Figure~\ref{fig:overview}), this means the current design of \vess methods "overfit" certain data distribution and metric. Thus, this literature may eagerly expect the emergence of "cross-node" \vess methods. Although some research efforts have proposed hybrid methods that combine graph and clustering techniques~\cite{chen2021spann}, such approaches are predominantly heuristic and lack concrete guidelines and adaptive mechanisms across diverse datasets. This highlights significant optimization opportunities, from automatically tuning index parameters and adaptively stitching methods to achieving fully dynamic and distribution-sensitive \vess mechanisms.

\section{Discussion}
\label{sec:discuss}

This section focuses on the lessons we have learned from \iceberg, as well as the current limitations that still remain.

\stitle{Lessons Learned.} We summarized three key lessons. \textbf{Lesson 1:}  Optimizing only for synthetic is insufficient since information is progressively lost across the \vess pipeline, and method improvements often fail to yield task-centric gains. End-to-end evaluation is essential to bridge the gap between academic benchmarks and real-world effectiveness. \textbf{Lesson 2:} Metric misuse and suboptimal method choices are major, yet avoidable, sources of information loss. A better understanding of embedding distributions can guide more effective alignment between models, metrics, and retrieval strategies.
\textbf{Lesson 3:} Current \vess methods remain relatively isolated in the end-to-end pipeline and lack dynamic optimization tailored to real-world scenarios.  There is untapped potential in developing {\em task-aware}, {\em metric-aware}, and {\em distribution}-aware retrieval strategies that dynamically adapt to real-world deployment needs.

\stitle{Limitation.} This study highlights hidden pitfalls in current retrieval methods and their implications for the future \vess, while leaving some aspects beyond its scope. (1) Euclidean distance and inner product (cosine as a special case) cover most needs while hyperbolic~\cite{ganea2018hyperbolic} and sparse embeddings introduce alternative metrics such as Mahalanobis distance, these remain niche directions. Given limited resources, we focus on representative tasks and will extend to such customized metrics as new needs arise . (2) It focuses on algorithmic characteristics rather than commercial system optimizations like cloud-native design or hardware heterogeneity. (3) Information loss from upstream machine learning is acknowledged as a critical issue, but no solution is proposed—it's flagged as a key area for future work. (4) The decision tree model in \iceberg is built on current \textsf{AI}-generated vectors, which may lead to "overfitting" as model landscapes evolve, necessitating ongoing updates, i.e., the meta-feature indicators and dividing thresholds may evolve with the literature development. As \textsf{AI} applications rapidly advance, future updates will incorporate emerging domains such as \textsf{LLM}-based agents, guided by the release of high-quality benchmark datasets.

\section{Conclusion}
\label{sec:conclude}

In this paper, we presented \textsf{Iceberg}, a comprehensive benchmark suite explicitly designed for end-to-end evaluation of vector similarity search (\vess) from task-centric perspectives, which shifts the focus from synthetic evaluation to the downstream task performance. Our study identifies the information loss funnel as the primary bottleneck in \vess and systematically analyzes its underlying causes through extensive experimentation. For empirical validation, \textsf{Iceberg} integrates eight large-scale datasets across four major domains, eight domain-specific representation models, thirteen state-of-the-art \vess methods, four downstream task metrics, and a variety of variations. To assist practitioners, it further provides an explainable decision-tree model that guides practitioners in selecting task-appropriate \vess methods. Finally, we summarize our findings and outline directions for future research, aimed at closing the gap between academic research and production deployments.

\bibliographystyle{ACM-Reference-Format}
\balance
\bibliography{ref}


\begin{thebibliography}{72}

%%% the Web address.

\ifx \showCODEN    \undefined \def \showCODEN     #1{\unskip}     \fi
\ifx \showISBNx    \undefined \def \showISBNx     #1{\unskip}     \fi
\ifx \showISBNxiii \undefined \def \showISBNxiii  #1{\unskip}     \fi
\ifx \showISSN     \undefined \def \showISSN      #1{\unskip}     \fi
\ifx \showLCCN     \undefined \def \showLCCN      #1{\unskip}     \fi
\ifx \shownote     \undefined \def \shownote      #1{#1}          \fi
\ifx \showarticletitle \undefined \def \showarticletitle #1{#1}   \fi
\ifx \showURL      \undefined \def \showURL       {\relax}        \fi
% invisible to TeX
\providecommand\bibfield[2]{#2}
\providecommand\bibinfo[2]{#2}
\providecommand\natexlab[1]{#1}
\providecommand\showeprint[2][]{arXiv:#2}

\bibitem[ima(2012)]%
        {imagenet_challenge}
 \bibinfo{year}{2012}\natexlab{}.
\newblock \bibinfo{howpublished}{\url{https://image-net.org/challenges/LSVRC/}}.
\newblock
\newblock
\shownote{Accessed: July 16, 2025}.


\bibitem[vit(2018)]%
        {vit_tiny_patch8_112}
 \bibinfo{year}{2018}\natexlab{}.
\newblock \bibinfo{howpublished}{\url{https://huggingface.co/gaunernst/vit_tiny_patch8_112.arcface_ms1mv3}}.
\newblock
\newblock
\shownote{Accessed: July 16, 2025}.


\bibitem[cvl(2018)]%
        {cvlface_arcface_ir101}
 \bibinfo{year}{2018}\natexlab{}.
\newblock \bibinfo{howpublished}{\url{https://huggingface.co/minchul/cvlface_arcface_ir101_webface4m}}.
\newblock
\newblock
\shownote{Accessed: July 16, 2025}.


\bibitem[con(2022)]%
        {convnext_large_mlp}
 \bibinfo{year}{2022}\natexlab{}.
\newblock \bibinfo{howpublished}{\url{https://huggingface.co/timm/convnext_large_mlp.clip_laion2b_soup_ft_in12k_in1k_384}}.
\newblock
\newblock
\shownote{Accessed: July 16, 2025}.


\bibitem[din(2023)]%
        {dinov2_base}
 \bibinfo{year}{2023}\natexlab{}.
\newblock \bibinfo{howpublished}{\url{https://huggingface.co/facebook/dinov2-base}}.
\newblock
\newblock
\shownote{Accessed: July 16, 2025}.


\bibitem[eva(2023)]%
        {eva02_large_patch14}
 \bibinfo{year}{2023}\natexlab{}.
\newblock \bibinfo{howpublished}{\url{https://huggingface.co/timm/eva02_large_patch14_448.mim_m38m_ft_in22k_in1k}}.
\newblock
\newblock
\shownote{Accessed: July 16, 2025}.


\bibitem[res(2024)]%
        {resflow_github}
 \bibinfo{year}{2024}\natexlab{}.
\newblock \bibinfo{howpublished}{\url{https://github.com/FuCongResearchSquad/ResFlow}}.
\newblock
\newblock
\shownote{Accessed: July 16, 2025}.


\bibitem[ste(2025)]%
        {stella_en_1_5B}
 \bibinfo{year}{2025}\natexlab{}.
\newblock \bibinfo{howpublished}{\url{https://huggingface.co/NovaSearch/stella_en_1.5B_v5}}.
\newblock
\newblock
\shownote{Accessed: July 16, 2025}.


\bibitem[Aguerrebere et~al\mbox{.}(2023)]%
        {aguerrebere2023similarity}
\bibfield{author}{\bibinfo{person}{Cecilia Aguerrebere}, \bibinfo{person}{Ishwar~Singh Bhati}, \bibinfo{person}{Mark Hildebrand}, \bibinfo{person}{Mariano Tepper}, {and} \bibinfo{person}{Theodore Willke}.} \bibinfo{year}{2023}\natexlab{}.
\newblock \showarticletitle{Similarity Search in the Blink of an Eye with Compressed Indices}.
\newblock \bibinfo{journal}{\emph{Proceedings of the VLDB Endowment}} \bibinfo{volume}{16}, \bibinfo{number}{11} (\bibinfo{year}{2023}), \bibinfo{pages}{3433--3446}.
\newblock
\newblock
\shownote{{\underline{\color{blue}\href{https://doi.org/10.14778/3611479.3611537} {https://doi.org/10.14778/3611479.3611537}}}}.


\bibitem[An et~al\mbox{.}(2021)]%
        {an2021partial}
\bibfield{author}{\bibinfo{person}{Xiang An}, \bibinfo{person}{Xuhan Zhu}, \bibinfo{person}{Yuan Gao}, \bibinfo{person}{Yang Xiao}, \bibinfo{person}{Yongle Zhao}, \bibinfo{person}{Ziyong Feng}, \bibinfo{person}{Lan Wu}, \bibinfo{person}{Bin Qin}, \bibinfo{person}{Ming Zhang}, \bibinfo{person}{Debing Zhang}, {et~al\mbox{.}}} \bibinfo{year}{2021}\natexlab{}.
\newblock \showarticletitle{Partial fc: Training 10 million identities on a single machine}. In \bibinfo{booktitle}{\emph{Proceedings of the IEEE/CVF International Conference on Computer Vision}}. \bibinfo{pages}{1445--1449}.
\newblock


\bibitem[Aum{\"u}ller et~al\mbox{.}(2020)]%
        {aumuller2020ann}
\bibfield{author}{\bibinfo{person}{Martin Aum{\"u}ller}, \bibinfo{person}{Erik Bernhardsson}, {and} \bibinfo{person}{Alexander Faithfull}.} \bibinfo{year}{2020}\natexlab{}.
\newblock \showarticletitle{ANN-Benchmarks: A benchmarking tool for approximate nearest neighbor algorithms}.
\newblock \bibinfo{journal}{\emph{Information Systems}}  \bibinfo{volume}{87} (\bibinfo{year}{2020}), \bibinfo{pages}{101374}.
\newblock
\newblock
\shownote{{\underline{\color{blue} \href{https://doi.org/10.1016/j.is.2019.02.006} {https://doi.org/10.1016/j.is.2019.02.006}}}}.


\bibitem[Azizi et~al\mbox{.}(2025)]%
        {azizi2025graph}
\bibfield{author}{\bibinfo{person}{Ilias Azizi}, \bibinfo{person}{Karima Echihabi}, {and} \bibinfo{person}{Themis Palpanas}.} \bibinfo{year}{2025}\natexlab{}.
\newblock \showarticletitle{Graph-Based Vector Search: An Experimental Evaluation of the State-of-the-Art}.
\newblock \bibinfo{journal}{\emph{Proceedings of the ACM on Management of Data}} \bibinfo{volume}{3}, \bibinfo{number}{1} (\bibinfo{year}{2025}), \bibinfo{pages}{1--31}.
\newblock
\newblock
\shownote{{\underline{\color{blue}\href{ https://doi.org/10.1145/3709693} { https://doi.org/10.1145/3709693}}}}.


\bibitem[Babenko and Lempitsky(2014)]%
        {babenko2014additive}
\bibfield{author}{\bibinfo{person}{Artem Babenko} {and} \bibinfo{person}{Victor Lempitsky}.} \bibinfo{year}{2014}\natexlab{}.
\newblock \showarticletitle{Additive quantization for extreme vector compression}. In \bibinfo{booktitle}{\emph{Proceedings of the IEEE Conference on Computer Vision and Pattern Recognition}}. \bibinfo{pages}{931--938}.
\newblock


\bibitem[Bai et~al\mbox{.}(2025)]%
        {bai2025qwen2}
\bibfield{author}{\bibinfo{person}{Shuai Bai}, \bibinfo{person}{Keqin Chen}, \bibinfo{person}{Xuejing Liu}, \bibinfo{person}{Jialin Wang}, \bibinfo{person}{Wenbin Ge}, \bibinfo{person}{Sibo Song}, \bibinfo{person}{Kai Dang}, \bibinfo{person}{Peng Wang}, \bibinfo{person}{Shijie Wang}, \bibinfo{person}{Jun Tang}, {et~al\mbox{.}}} \bibinfo{year}{2025}\natexlab{}.
\newblock \showarticletitle{Qwen2. 5-vl technical report}.
\newblock \bibinfo{journal}{\emph{arXiv preprint arXiv:2502.13923}} (\bibinfo{year}{2025}).
\newblock


\bibitem[Chen et~al\mbox{.}(2024)]%
        {chen2024roargraph}
\bibfield{author}{\bibinfo{person}{Meng Chen}, \bibinfo{person}{Kai Zhang}, \bibinfo{person}{Zhenying He}, \bibinfo{person}{Yinan Jing}, {and} \bibinfo{person}{X~Sean Wang}.} \bibinfo{year}{2024}\natexlab{}.
\newblock \showarticletitle{RoarGraph: A Projected Bipartite Graph for Efficient Cross-Modal Approximate Nearest Neighbor Search}.
\newblock \bibinfo{journal}{\emph{Proceedings of the VLDB Endowment}} \bibinfo{volume}{17}, \bibinfo{number}{11} (\bibinfo{year}{2024}), \bibinfo{pages}{2735--2749}.
\newblock
\newblock
\shownote{{\underline{\color{blue}\href{ https://doi.org/10.14778/3681954.368195} { https://doi.org/10.14778/3681954.368195}}}}.


\bibitem[Chen et~al\mbox{.}(2021)]%
        {chen2021spann}
\bibfield{author}{\bibinfo{person}{Qi Chen}, \bibinfo{person}{Bing Zhao}, \bibinfo{person}{Haidong Wang}, \bibinfo{person}{Mingqin Li}, \bibinfo{person}{Chuanjie Liu}, \bibinfo{person}{Zengzhong Li}, \bibinfo{person}{Mao Yang}, {and} \bibinfo{person}{Jingdong Wang}.} \bibinfo{year}{2021}\natexlab{}.
\newblock \showarticletitle{Spann: Highly-efficient billion-scale approximate nearest neighborhood search}.
\newblock \bibinfo{journal}{\emph{Advances in Neural Information Processing Systems}}  \bibinfo{volume}{34} (\bibinfo{year}{2021}), \bibinfo{pages}{5199--5212}.
\newblock


\bibitem[Chen et~al\mbox{.}(2025a)]%
        {chen2025stitching}
\bibfield{author}{\bibinfo{person}{Tingyang Chen}, \bibinfo{person}{Cong Fu}, \bibinfo{person}{Xiangyu Ke}, \bibinfo{person}{Yunjun Gao}, \bibinfo{person}{Yabo Ni}, {and} \bibinfo{person}{Anxiang Zeng}.} \bibinfo{year}{2025}\natexlab{a}.
\newblock \showarticletitle{Stitching Inner Product and Euclidean Metrics for Topology-aware Maximum Inner Product Search}.
\newblock \bibinfo{journal}{\emph{arXiv preprint arXiv:2504.14861}} (\bibinfo{year}{2025}).
\newblock


\bibitem[Chen et~al\mbox{.}(2025b)]%
        {chen2025maximum}
\bibfield{author}{\bibinfo{person}{Tingyang Chen}, \bibinfo{person}{Cong Fu}, \bibinfo{person}{Kun Wang}, \bibinfo{person}{Xiangyu Ke}, \bibinfo{person}{Yunjun Gao}, \bibinfo{person}{Wenchao Zhou}, \bibinfo{person}{Yabo Ni}, {and} \bibinfo{person}{Anxiang Zeng}.} \bibinfo{year}{2025}\natexlab{b}.
\newblock \showarticletitle{Maximum Inner Product is Query-Scaled Nearest Neighbor}.
\newblock \bibinfo{journal}{\emph{arXiv preprint arXiv:2503.06882}} (\bibinfo{year}{2025}).
\newblock


\bibitem[Chi et~al\mbox{.}(2022)]%
        {chi2022representation}
\bibfield{author}{\bibinfo{person}{Zewen Chi}, \bibinfo{person}{Li Dong}, \bibinfo{person}{Shaohan Huang}, \bibinfo{person}{Damai Dai}, \bibinfo{person}{Shuming Ma}, \bibinfo{person}{Barun Patra}, \bibinfo{person}{Saksham Singhal}, \bibinfo{person}{Payal Bajaj}, \bibinfo{person}{Xia Song}, \bibinfo{person}{Xian-Ling Mao}, {et~al\mbox{.}}} \bibinfo{year}{2022}\natexlab{}.
\newblock \showarticletitle{On the representation collapse of sparse mixture of experts}.
\newblock \bibinfo{journal}{\emph{Advances in Neural Information Processing Systems}}  \bibinfo{volume}{35} (\bibinfo{year}{2022}), \bibinfo{pages}{34600--34613}.
\newblock


\bibitem[Costa et~al\mbox{.}(2005)]%
        {costa2005estimating}
\bibfield{author}{\bibinfo{person}{Jose~A Costa}, \bibinfo{person}{Abhishek Girotra}, {and} \bibinfo{person}{AO Hero}.} \bibinfo{year}{2005}\natexlab{}.
\newblock \showarticletitle{Estimating local intrinsic dimension with k-nearest neighbor graphs}. In \bibinfo{booktitle}{\emph{IEEE/SP 13th Workshop on Statistical Signal Processing, 2005}}. \bibinfo{pages}{417--422}.
\newblock


\bibitem[Datar et~al\mbox{.}(2004)]%
        {datar2004locality}
\bibfield{author}{\bibinfo{person}{Mayur Datar}, \bibinfo{person}{Nicole Immorlica}, \bibinfo{person}{Piotr Indyk}, {and} \bibinfo{person}{Vahab~S Mirrokni}.} \bibinfo{year}{2004}\natexlab{}.
\newblock \showarticletitle{Locality-sensitive hashing scheme based on p-stable distributions}. In \bibinfo{booktitle}{\emph{Proceedings of the twentieth annual symposium on Computational geometry}}. \bibinfo{pages}{253--262}.
\newblock


\bibitem[Deng et~al\mbox{.}(2019)]%
        {deng2019arcface}
\bibfield{author}{\bibinfo{person}{Jiankang Deng}, \bibinfo{person}{Jia Guo}, \bibinfo{person}{Niannan Xue}, {and} \bibinfo{person}{Stefanos Zafeiriou}.} \bibinfo{year}{2019}\natexlab{}.
\newblock \showarticletitle{Arcface: Additive angular margin loss for deep face recognition}. In \bibinfo{booktitle}{\emph{Proceedings of the IEEE/CVF conference on computer vision and pattern recognition}}. \bibinfo{pages}{4690--4699}.
\newblock


\bibitem[Douze et~al\mbox{.}(2024)]%
        {douze2024faiss}
\bibfield{author}{\bibinfo{person}{Matthijs Douze}, \bibinfo{person}{Alexandr Guzhva}, \bibinfo{person}{Chengqi Deng}, \bibinfo{person}{Jeff Johnson}, \bibinfo{person}{Gergely Szilvasy}, \bibinfo{person}{Pierre-Emmanuel Mazar{\'e}}, \bibinfo{person}{Maria Lomeli}, \bibinfo{person}{Lucas Hosseini}, {and} \bibinfo{person}{Herv{\'e} J{\'e}gou}.} \bibinfo{year}{2024}\natexlab{}.
\newblock \showarticletitle{The faiss library}.
\newblock \bibinfo{journal}{\emph{arXiv preprint arXiv:2401.08281}} (\bibinfo{year}{2024}).
\newblock


\bibitem[Fang et~al\mbox{.}(2024)]%
        {fang2024eva}
\bibfield{author}{\bibinfo{person}{Yuxin Fang}, \bibinfo{person}{Quan Sun}, \bibinfo{person}{Xinggang Wang}, \bibinfo{person}{Tiejun Huang}, \bibinfo{person}{Xinlong Wang}, {and} \bibinfo{person}{Yue Cao}.} \bibinfo{year}{2024}\natexlab{}.
\newblock \showarticletitle{Eva-02: A visual representation for neon genesis}.
\newblock \bibinfo{journal}{\emph{Image and Vision Computing}}  \bibinfo{volume}{149} (\bibinfo{year}{2024}), \bibinfo{pages}{105171}.
\newblock
\newblock
\shownote{{\underline{\color{blue}\href{https://doi.org/10.1016/j.imavis.2024.105171} {https://doi.org/10.1016/j.imavis.2024.105171}}}}.


\bibitem[Ferhatosmanoglu et~al\mbox{.}(2000)]%
        {ferhatosmanoglu2000vector}
\bibfield{author}{\bibinfo{person}{Hakan Ferhatosmanoglu}, \bibinfo{person}{Ertem Tuncel}, \bibinfo{person}{Divyakant Agrawal}, {and} \bibinfo{person}{Amr El~Abbadi}.} \bibinfo{year}{2000}\natexlab{}.
\newblock \showarticletitle{Vector approximation based indexing for non-uniform high dimensional data sets}. In \bibinfo{booktitle}{\emph{Proceedings of the ninth international conference on Information and knowledge management}}. \bibinfo{pages}{202--209}.
\newblock


\bibitem[Fu et~al\mbox{.}(2024)]%
        {fu2024residual}
\bibfield{author}{\bibinfo{person}{Cong Fu}, \bibinfo{person}{Kun Wang}, \bibinfo{person}{Jiahua Wu}, \bibinfo{person}{Yizhou Chen}, \bibinfo{person}{Guangda Huzhang}, \bibinfo{person}{Yabo Ni}, \bibinfo{person}{Anxiang Zeng}, {and} \bibinfo{person}{Zhiming Zhou}.} \bibinfo{year}{2024}\natexlab{}.
\newblock \showarticletitle{Residual Multi-Task Learner for Applied Ranking}. In \bibinfo{booktitle}{\emph{Proceedings of the 30th ACM SIGKDD conference on knowledge discovery and data Mining}}. \bibinfo{pages}{4974--4985}.
\newblock


\bibitem[Fu et~al\mbox{.}(2019)]%
        {fu2019fast}
\bibfield{author}{\bibinfo{person}{Cong Fu}, \bibinfo{person}{Chao Xiang}, \bibinfo{person}{Changxu Wang}, {and} \bibinfo{person}{Deng Cai}.} \bibinfo{year}{2019}\natexlab{}.
\newblock \showarticletitle{Fast approximate nearest neighbor search with the navigating spreading-out graph}.
\newblock \bibinfo{journal}{\emph{Proceedings of the VLDB Endowment}} \bibinfo{volume}{12}, \bibinfo{number}{5} (\bibinfo{year}{2019}), \bibinfo{pages}{461--474}.
\newblock
\newblock
\shownote{{\underline{\color{blue}\href{https://doi.org/10.14778/3303753.3303754} {https://doi.org/10.14778/3303753.3303754}}}}.


\bibitem[Ganea et~al\mbox{.}(2018)]%
        {ganea2018hyperbolic}
\bibfield{author}{\bibinfo{person}{Octavian Ganea}, \bibinfo{person}{Gary B{\'e}cigneul}, {and} \bibinfo{person}{Thomas Hofmann}.} \bibinfo{year}{2018}\natexlab{}.
\newblock \showarticletitle{Hyperbolic neural networks}.
\newblock \bibinfo{journal}{\emph{NeurIPS}}  \bibinfo{volume}{31} (\bibinfo{year}{2018}).
\newblock


\bibitem[Gao et~al\mbox{.}(2025a)]%
        {gaohigh}
\bibfield{author}{\bibinfo{person}{Jianyang Gao}, \bibinfo{person}{Yutong Gou}, \bibinfo{person}{Yuexuan Xu}, \bibinfo{person}{Jifan Shi}, \bibinfo{person}{Cheng Long}, \bibinfo{person}{Raymond Chi-Wing Wong}, {and} \bibinfo{person}{Themis Palpanas}.} \bibinfo{year}{2025}\natexlab{a}.
\newblock \showarticletitle{High-Dimensional Vector Quantization: General Framework, Recent Advances, and Future Directions}.
\newblock \bibinfo{journal}{\emph{Data Engineering}} (\bibinfo{year}{2025}), \bibinfo{pages}{3}.
\newblock


\bibitem[Gao et~al\mbox{.}(2025b)]%
        {gao2025practical}
\bibfield{author}{\bibinfo{person}{Jianyang Gao}, \bibinfo{person}{Yutong Gou}, \bibinfo{person}{Yuexuan Xu}, \bibinfo{person}{Yongyi Yang}, \bibinfo{person}{Cheng Long}, {and} \bibinfo{person}{Raymond Chi-Wing Wong}.} \bibinfo{year}{2025}\natexlab{b}.
\newblock \showarticletitle{Practical and asymptotically optimal quantization of high-dimensional vectors in euclidean space for approximate nearest neighbor search}.
\newblock \bibinfo{journal}{\emph{Proceedings of the ACM on Management of Data}} \bibinfo{volume}{3}, \bibinfo{number}{3} (\bibinfo{year}{2025}), \bibinfo{pages}{1--26}.
\newblock
\newblock
\shownote{{\underline{\color{blue}\href{https://doi.org/10.1145/3725413} {https://doi.org/10.1145/3725413}}}}.


\bibitem[Gao and Long(2024)]%
        {gao2024rabitq}
\bibfield{author}{\bibinfo{person}{Jianyang Gao} {and} \bibinfo{person}{Cheng Long}.} \bibinfo{year}{2024}\natexlab{}.
\newblock \showarticletitle{RaBitQ: quantizing high-dimensional vectors with a theoretical error bound for approximate nearest neighbor search}.
\newblock \bibinfo{journal}{\emph{Proceedings of the ACM on Management of Data}} \bibinfo{volume}{2}, \bibinfo{number}{3} (\bibinfo{year}{2024}), \bibinfo{pages}{1--27}.
\newblock
\newblock
\shownote{{\underline{\color{blue}\href{https://doi.org/10.1145/3654970} {https://doi.org/10.1145/3654970}}}}.


\bibitem[Gao et~al\mbox{.}(2018)]%
        {gao2018recommendation}
\bibfield{author}{\bibinfo{person}{Li Gao}, \bibinfo{person}{Hong Yang}, \bibinfo{person}{Jia Wu}, \bibinfo{person}{Chuan Zhou}, \bibinfo{person}{Weixue Lu}, {and} \bibinfo{person}{Yue Hu}.} \bibinfo{year}{2018}\natexlab{}.
\newblock \showarticletitle{Recommendation with multi-source heterogeneous information}. In \bibinfo{booktitle}{\emph{Proceedings of the 27th International Joint Conference on Artificial Intelligence}}. \bibinfo{pages}{3378--3384}.
\newblock


\bibitem[Guo et~al\mbox{.}(2020)]%
        {guo2020accelerating}
\bibfield{author}{\bibinfo{person}{Ruiqi Guo}, \bibinfo{person}{Philip Sun}, \bibinfo{person}{Erik Lindgren}, \bibinfo{person}{Quan Geng}, \bibinfo{person}{David Simcha}, \bibinfo{person}{Felix Chern}, {and} \bibinfo{person}{Sanjiv Kumar}.} \bibinfo{year}{2020}\natexlab{}.
\newblock \showarticletitle{Accelerating large-scale inference with anisotropic vector quantization}. In \bibinfo{booktitle}{\emph{International Conference on Machine Learning}}. \bibinfo{pages}{3887--3896}.
\newblock


\bibitem[Hsieh et~al\mbox{.}(2017)]%
        {hsieh2017collaborative}
\bibfield{author}{\bibinfo{person}{Cheng-Kang Hsieh}, \bibinfo{person}{Longqi Yang}, \bibinfo{person}{Yin Cui}, \bibinfo{person}{Tsung-Yi Lin}, \bibinfo{person}{Serge Belongie}, {and} \bibinfo{person}{Deborah Estrin}.} \bibinfo{year}{2017}\natexlab{}.
\newblock \showarticletitle{Collaborative metric learning}. In \bibinfo{booktitle}{\emph{Proceedings of the 26th international conference on world wide web}}. \bibinfo{pages}{193--201}.
\newblock


\bibitem[Huang et~al\mbox{.}(2022)]%
        {huang2022riemannian}
\bibfield{author}{\bibinfo{person}{Chin-Wei Huang}, \bibinfo{person}{Milad Aghajohari}, \bibinfo{person}{Joey Bose}, \bibinfo{person}{Prakash Panangaden}, {and} \bibinfo{person}{Aaron~C Courville}.} \bibinfo{year}{2022}\natexlab{}.
\newblock \showarticletitle{Riemannian diffusion models}.
\newblock \bibinfo{journal}{\emph{Advances in Neural Information Processing Systems}}  \bibinfo{volume}{35} (\bibinfo{year}{2022}), \bibinfo{pages}{2750--2761}.
\newblock


\bibitem[Huang et~al\mbox{.}(2015)]%
        {huang2015query}
\bibfield{author}{\bibinfo{person}{Qiang Huang}, \bibinfo{person}{Jianlin Feng}, \bibinfo{person}{Yikai Zhang}, \bibinfo{person}{Qiong Fang}, {and} \bibinfo{person}{Wilfred Ng}.} \bibinfo{year}{2015}\natexlab{}.
\newblock \showarticletitle{Query-aware locality-sensitive hashing for approximate nearest neighbor search}.
\newblock \bibinfo{journal}{\emph{Proceedings of the VLDB Endowment}} \bibinfo{volume}{9}, \bibinfo{number}{1} (\bibinfo{year}{2015}), \bibinfo{pages}{1--12}.
\newblock


\bibitem[Jayaram~Subramanya et~al\mbox{.}(2019)]%
        {jayaram2019diskann}
\bibfield{author}{\bibinfo{person}{Suhas Jayaram~Subramanya}, \bibinfo{person}{Fnu Devvrit}, \bibinfo{person}{Harsha~Vardhan Simhadri}, \bibinfo{person}{Ravishankar Krishnawamy}, {and} \bibinfo{person}{Rohan Kadekodi}.} \bibinfo{year}{2019}\natexlab{}.
\newblock \showarticletitle{Diskann: Fast accurate billion-point nearest neighbor search on a single node}.
\newblock \bibinfo{journal}{\emph{Advances in neural information processing Systems}}  \bibinfo{volume}{32} (\bibinfo{year}{2019}).
\newblock


\bibitem[Kang et~al\mbox{.}({[n.\,d.]})]%
        {kangbigvectorbench}
\bibfield{author}{\bibinfo{person}{Guoxin Kang}, \bibinfo{person}{Zhongxin Ge}, \bibinfo{person}{Jingpei Hu}, \bibinfo{person}{Xueya Zhang}, \bibinfo{person}{Lei Wang}, {and} \bibinfo{person}{Jianfeng Zhan}.} \bibinfo{year}{[n.\,d.]}\natexlab{}.
\newblock \showarticletitle{BigVectorBench: Heterogeneous Data Embedding and Compound eries are Essential in Evaluating Vector Databases}.
\newblock  (\bibinfo{year}{[n.\,d.]}).
\newblock


\bibitem[Koenigstein et~al\mbox{.}(2012)]%
        {koenigstein2012efficient}
\bibfield{author}{\bibinfo{person}{Noam Koenigstein}, \bibinfo{person}{Parikshit Ram}, {and} \bibinfo{person}{Yuval Shavitt}.} \bibinfo{year}{2012}\natexlab{}.
\newblock \showarticletitle{Efficient retrieval of recommendations in a matrix factorization framework}. In \bibinfo{booktitle}{\emph{Proceedings of the 21st ACM international conference on Information and knowledge management}}. \bibinfo{pages}{535--544}.
\newblock


\bibitem[Krizhevsky et~al\mbox{.}(2012)]%
        {krizhevsky2012imagenet}
\bibfield{author}{\bibinfo{person}{Alex Krizhevsky}, \bibinfo{person}{Ilya Sutskever}, {and} \bibinfo{person}{Geoffrey~E Hinton}.} \bibinfo{year}{2012}\natexlab{}.
\newblock \showarticletitle{Imagenet classification with deep convolutional neural networks}.
\newblock \bibinfo{journal}{\emph{Advances in neural information processing systems}}  \bibinfo{volume}{25} (\bibinfo{year}{2012}).
\newblock


\bibitem[Kulis et~al\mbox{.}(2013)]%
        {kulis2013metric}
\bibfield{author}{\bibinfo{person}{Brian Kulis} {et~al\mbox{.}}} \bibinfo{year}{2013}\natexlab{}.
\newblock \showarticletitle{Metric learning: A survey}.
\newblock \bibinfo{journal}{\emph{Foundations and Trends{\textregistered} in Machine Learning}} \bibinfo{volume}{5}, \bibinfo{number}{4} (\bibinfo{year}{2013}), \bibinfo{pages}{287--364}.
\newblock


\bibitem[Lewis et~al\mbox{.}(2020)]%
        {lewis2020retrieval}
\bibfield{author}{\bibinfo{person}{Patrick Lewis}, \bibinfo{person}{Ethan Perez}, \bibinfo{person}{Aleksandra Piktus}, \bibinfo{person}{Fabio Petroni}, \bibinfo{person}{Vladimir Karpukhin}, \bibinfo{person}{Naman Goyal}, \bibinfo{person}{Heinrich K{\"u}ttler}, \bibinfo{person}{Mike Lewis}, \bibinfo{person}{Wen-tau Yih}, \bibinfo{person}{Tim Rockt{\"a}schel}, {et~al\mbox{.}}} \bibinfo{year}{2020}\natexlab{}.
\newblock \showarticletitle{Retrieval-augmented generation for knowledge-intensive nlp tasks}.
\newblock \bibinfo{journal}{\emph{Advances in Neural Information Processing Systems}}  \bibinfo{volume}{33} (\bibinfo{year}{2020}), \bibinfo{pages}{9459--9474}.
\newblock


\bibitem[Li et~al\mbox{.}(2022)]%
        {li2022understanding}
\bibfield{author}{\bibinfo{person}{Alexander~C Li}, \bibinfo{person}{Alexei~A Efros}, {and} \bibinfo{person}{Deepak Pathak}.} \bibinfo{year}{2022}\natexlab{}.
\newblock \showarticletitle{Understanding collapse in non-contrastive siamese representation learning}. In \bibinfo{booktitle}{\emph{European Conference on Computer Vision}}. \bibinfo{pages}{490--505}.
\newblock


\bibitem[Li et~al\mbox{.}(2019)]%
        {li2019approximate}
\bibfield{author}{\bibinfo{person}{Wen Li}, \bibinfo{person}{Ying Zhang}, \bibinfo{person}{Yifang Sun}, \bibinfo{person}{Wei Wang}, \bibinfo{person}{Mingjie Li}, \bibinfo{person}{Wenjie Zhang}, {and} \bibinfo{person}{Xuemin Lin}.} \bibinfo{year}{2019}\natexlab{}.
\newblock \showarticletitle{Approximate nearest neighbor search on high dimensional data—experiments, analyses, and improvement}.
\newblock \bibinfo{journal}{\emph{IEEE Transactions on Knowledge and Data Engineering}} \bibinfo{volume}{32}, \bibinfo{number}{8} (\bibinfo{year}{2019}), \bibinfo{pages}{1475--1488}.
\newblock
\newblock
\shownote{{\underline{\color{blue} \href{https://doi.org/10.1109/TKDE.2019.2909204} {https://doi.org/10.1109/TKDE.2019.2909204}}}}.


\bibitem[Li et~al\mbox{.}(2024)]%
        {li2024nlp}
\bibfield{author}{\bibinfo{person}{Yuangang Li}, \bibinfo{person}{Jiaqi Li}, \bibinfo{person}{Zhuo Xiao}, \bibinfo{person}{Tiankai Yang}, \bibinfo{person}{Yi Nian}, \bibinfo{person}{Xiyang Hu}, {and} \bibinfo{person}{Yue Zhao}.} \bibinfo{year}{2024}\natexlab{}.
\newblock \showarticletitle{Nlp-adbench: Nlp anomaly detection benchmark}.
\newblock \bibinfo{journal}{\emph{arXiv preprint arXiv:2412.04784}} (\bibinfo{year}{2024}).
\newblock


\bibitem[Liu et~al\mbox{.}(2020)]%
        {liu2020understanding}
\bibfield{author}{\bibinfo{person}{Jie Liu}, \bibinfo{person}{Xiao Yan}, \bibinfo{person}{Xinyan Dai}, \bibinfo{person}{Zhirong Li}, \bibinfo{person}{James Cheng}, {and} \bibinfo{person}{Ming-Chang Yang}.} \bibinfo{year}{2020}\natexlab{}.
\newblock \showarticletitle{Understanding and improving proximity graph based maximum inner product search}. In \bibinfo{booktitle}{\emph{Proceedings of the AAAI Conference on Artificial Intelligence}}, Vol.~\bibinfo{volume}{34}. \bibinfo{pages}{139--146}.
\newblock


\bibitem[Liu et~al\mbox{.}(2014)]%
        {liu2014sk}
\bibfield{author}{\bibinfo{person}{Yingfan Liu}, \bibinfo{person}{Jiangtao Cui}, \bibinfo{person}{Zi Huang}, \bibinfo{person}{Hui Li}, {and} \bibinfo{person}{Heng~Tao Shen}.} \bibinfo{year}{2014}\natexlab{}.
\newblock \showarticletitle{SK-LSH: an efficient index structure for approximate nearest neighbor search}.
\newblock \bibinfo{journal}{\emph{Proceedings of the VLDB Endowment}} \bibinfo{volume}{7}, \bibinfo{number}{9} (\bibinfo{year}{2014}), \bibinfo{pages}{745--756}.
\newblock


\bibitem[Lowe(2004)]%
        {lowe2004distinctive}
\bibfield{author}{\bibinfo{person}{David~G Lowe}.} \bibinfo{year}{2004}\natexlab{}.
\newblock \showarticletitle{Distinctive image features from scale-invariant keypoints}.
\newblock \bibinfo{journal}{\emph{International journal of computer vision}}  \bibinfo{volume}{60} (\bibinfo{year}{2004}), \bibinfo{pages}{91--110}.
\newblock


\bibitem[Lu et~al\mbox{.}(2021)]%
        {lu2021hvs}
\bibfield{author}{\bibinfo{person}{Kejing Lu}, \bibinfo{person}{Mineichi Kudo}, \bibinfo{person}{Chuan Xiao}, {and} \bibinfo{person}{Yoshiharu Ishikawa}.} \bibinfo{year}{2021}\natexlab{}.
\newblock \showarticletitle{HVS: hierarchical graph structure based on voronoi diagrams for solving approximate nearest neighbor search}.
\newblock \bibinfo{journal}{\emph{Proceedings of the VLDB Endowment}} \bibinfo{volume}{15}, \bibinfo{number}{2} (\bibinfo{year}{2021}), \bibinfo{pages}{246--258}.
\newblock
\newblock
\shownote{{\underline{\color{blue}\href{ https://doi.org/10.14778/3489496.3489506} { https://doi.org/10.14778/3489496.3489506}}}}.


\bibitem[Ma et~al\mbox{.}(2024)]%
        {ma2024reconsider}
\bibfield{author}{\bibinfo{person}{Hengzhao Ma}, \bibinfo{person}{Jianzhong Li}, {and} \bibinfo{person}{Yong Zhang}.} \bibinfo{year}{2024}\natexlab{}.
\newblock \showarticletitle{Reconsidering Tree based Methods for k-Maximum Inner-Product Search: The LRUS-CoverTree}. In \bibinfo{booktitle}{\emph{2024 IEEE 40th international conference on data engineering}}.
\newblock


\bibitem[Malkov and Yashunin(2018)]%
        {malkov2018efficient}
\bibfield{author}{\bibinfo{person}{Yu~A Malkov} {and} \bibinfo{person}{Dmitry~A Yashunin}.} \bibinfo{year}{2018}\natexlab{}.
\newblock \showarticletitle{Efficient and robust approximate nearest neighbor search using hierarchical navigable small world graphs}.
\newblock \bibinfo{journal}{\emph{IEEE transactions on pattern analysis and machine intelligence}} \bibinfo{volume}{42}, \bibinfo{number}{4} (\bibinfo{year}{2018}), \bibinfo{pages}{824--836}.
\newblock
\newblock
\shownote{{\underline{\color{blue}\href{https://doi.org/10.1109/TPAMI.2018.2889473} {https://doi.org/10.1109/TPAMI.2018.2889473}}}}.


\bibitem[Matsui et~al\mbox{.}(2018)]%
        {matsui2018survey}
\bibfield{author}{\bibinfo{person}{Yusuke Matsui}, \bibinfo{person}{Yusuke Uchida}, \bibinfo{person}{Herv{\'e} J{\'e}gou}, {and} \bibinfo{person}{Shin'ichi Satoh}.} \bibinfo{year}{2018}\natexlab{}.
\newblock \showarticletitle{A survey of product quantization}.
\newblock \bibinfo{journal}{\emph{ITE Transactions on Media Technology and Applications}} \bibinfo{volume}{6}, \bibinfo{number}{1} (\bibinfo{year}{2018}), \bibinfo{pages}{2--10}.
\newblock
\newblock
\shownote{{\underline{\color{blue}\href{ https://doi.org/10.3169/mta.6.2} { https://doi.org/10.3169/mta.6.2}}}}.


\bibitem[Morozov and Babenko(2018)]%
        {morozov2018non}
\bibfield{author}{\bibinfo{person}{Stanislav Morozov} {and} \bibinfo{person}{Artem Babenko}.} \bibinfo{year}{2018}\natexlab{}.
\newblock \showarticletitle{Non-metric similarity graphs for maximum inner product search}.
\newblock \bibinfo{journal}{\emph{Advances in Neural Information Processing Systems}}  \bibinfo{volume}{31} (\bibinfo{year}{2018}).
\newblock


\bibitem[Muennighoff et~al\mbox{.}(2023)]%
        {muennighoff2023mteb}
\bibfield{author}{\bibinfo{person}{Niklas Muennighoff}, \bibinfo{person}{Nouamane Tazi}, \bibinfo{person}{Loic Magne}, {and} \bibinfo{person}{Nils Reimers}.} \bibinfo{year}{2023}\natexlab{}.
\newblock \showarticletitle{MTEB: Massive Text Embedding Benchmark}. In \bibinfo{booktitle}{\emph{Proceedings of the 17th Conference of the European Chapter of the Association for Computational Linguistics}}. \bibinfo{pages}{2014--2037}.
\newblock


\bibitem[Oquab et~al\mbox{.}(2024)]%
        {oquab2024dinov2}
\bibfield{author}{\bibinfo{person}{Maxime Oquab}, \bibinfo{person}{Timoth{\'e}e Darcet}, \bibinfo{person}{Th{\'e}o Moutakanni}, \bibinfo{person}{Huy Vo}, \bibinfo{person}{Marc Szafraniec}, \bibinfo{person}{Vasil Khalidov}, \bibinfo{person}{Pierre Fernandez}, \bibinfo{person}{Daniel Haziza}, \bibinfo{person}{Francisco Massa}, \bibinfo{person}{Alaaeldin El-Nouby}, {et~al\mbox{.}}} \bibinfo{year}{2024}\natexlab{}.
\newblock \showarticletitle{DINOv2: Learning Robust Visual Features without Supervision}.
\newblock \bibinfo{journal}{\emph{Transactions on Machine Learning Research Journal}} (\bibinfo{year}{2024}), \bibinfo{pages}{1--31}.
\newblock
\newblock
\shownote{{\underline{\color{blue} \href{https://dx.doi.org/10.48550/arxiv.2304.07193} {https://dx.doi.org/10.48550/arxiv.2304.07193}}}}.


\bibitem[Pan et~al\mbox{.}(2024)]%
        {pan2024survey}
\bibfield{author}{\bibinfo{person}{James~Jie Pan}, \bibinfo{person}{Jianguo Wang}, {and} \bibinfo{person}{Guoliang Li}.} \bibinfo{year}{2024}\natexlab{}.
\newblock \showarticletitle{Survey of vector database management systems}.
\newblock \bibinfo{journal}{\emph{The VLDB Journal}} \bibinfo{volume}{33}, \bibinfo{number}{5} (\bibinfo{year}{2024}), \bibinfo{pages}{1591--1615}.
\newblock
\newblock
\shownote{{\underline{\color{blue} \href{https://doi.org/10.1007/s00778-024-00864-x} {https://doi.org/10.1007/s00778-024-00864-x}}}}.


\bibitem[Papyan et~al\mbox{.}(2020)]%
        {papyan2020prevalence}
\bibfield{author}{\bibinfo{person}{Vardan Papyan}, \bibinfo{person}{XY Han}, {and} \bibinfo{person}{David~L Donoho}.} \bibinfo{year}{2020}\natexlab{}.
\newblock \showarticletitle{Prevalence of neural collapse during the terminal phase of deep learning training}.
\newblock \bibinfo{journal}{\emph{Proceedings of the National Academy of Sciences}} \bibinfo{volume}{117}, \bibinfo{number}{40} (\bibinfo{year}{2020}), \bibinfo{pages}{24652--24663}.
\newblock


\bibitem[Peng et~al\mbox{.}(2023)]%
        {peng2023efficient}
\bibfield{author}{\bibinfo{person}{Yun Peng}, \bibinfo{person}{Byron Choi}, \bibinfo{person}{Tsz~Nam Chan}, \bibinfo{person}{Jianye Yang}, {and} \bibinfo{person}{Jianliang Xu}.} \bibinfo{year}{2023}\natexlab{}.
\newblock \showarticletitle{Efficient approximate nearest neighbor search in multi-dimensional databases}.
\newblock \bibinfo{journal}{\emph{Proceedings of the ACM on Management of Data}} \bibinfo{volume}{1}, \bibinfo{number}{1} (\bibinfo{year}{2023}), \bibinfo{pages}{1--27}.
\newblock
\newblock
\shownote{{\underline{\color{blue}\href{https://doi.org/10.1145/3588908} {https://doi.org/10.1145/3588908}}}}.


\bibitem[Ram and Gray(2012)]%
        {ram2012maximum}
\bibfield{author}{\bibinfo{person}{Parikshit Ram} {and} \bibinfo{person}{Alexander~G Gray}.} \bibinfo{year}{2012}\natexlab{}.
\newblock \showarticletitle{Maximum inner-product search using cone trees}. In \bibinfo{booktitle}{\emph{Proceedings of the 18th ACM SIGKDD international conference on Knowledge discovery and data mining}}. \bibinfo{pages}{931--939}.
\newblock


\bibitem[Ram and Sinha(2019)]%
        {ram2019revisiting}
\bibfield{author}{\bibinfo{person}{Parikshit Ram} {and} \bibinfo{person}{Kaushik Sinha}.} \bibinfo{year}{2019}\natexlab{}.
\newblock \showarticletitle{Revisiting kd-tree for nearest neighbor search}. In \bibinfo{booktitle}{\emph{Proceedings of the 25th ACM SIGKDD conference on knowledge discovery and data mining}}. \bibinfo{pages}{1378--1388}.
\newblock


\bibitem[Ruan et~al\mbox{.}(2025)]%
        {ruan2025empowering}
\bibfield{author}{\bibinfo{person}{Jiancheng Ruan}, \bibinfo{person}{Tingyang Chen}, \bibinfo{person}{Renchi Yang}, \bibinfo{person}{Xiangyu Ke}, {and} \bibinfo{person}{Yunjun Gao}.} \bibinfo{year}{2025}\natexlab{}.
\newblock \showarticletitle{Empowering Graph-based Approximate Nearest Neighbor Search with Adaptive Awareness Capabilities}.
\newblock \bibinfo{journal}{\emph{arXiv preprint arXiv:2506.15986}} (\bibinfo{year}{2025}).
\newblock


\bibitem[Safavi and Koutra(2020)]%
        {safavi2020codex}
\bibfield{author}{\bibinfo{person}{Tara Safavi} {and} \bibinfo{person}{Danai Koutra}.} \bibinfo{year}{2020}\natexlab{}.
\newblock \showarticletitle{Codex: A comprehensive knowledge graph completion benchmark}.
\newblock \bibinfo{journal}{\emph{arXiv preprint arXiv:2009.07810}} (\bibinfo{year}{2020}).
\newblock


\bibitem[Shrivastava and Li(2014)]%
        {shrivastava2014asymmetric}
\bibfield{author}{\bibinfo{person}{Anshumali Shrivastava} {and} \bibinfo{person}{Ping Li}.} \bibinfo{year}{2014}\natexlab{}.
\newblock \showarticletitle{Asymmetric LSH (ALSH) for sublinear time maximum inner product search (MIPS)}. In \bibinfo{booktitle}{\emph{Advances in neural information processing systems}}. \bibinfo{pages}{2321--2329}.
\newblock


\bibitem[Simhadri et~al\mbox{.}(2022)]%
        {simhadri2022results}
\bibfield{author}{\bibinfo{person}{Harsha~Vardhan Simhadri}, \bibinfo{person}{George Williams}, \bibinfo{person}{Martin Aum{\"u}ller}, \bibinfo{person}{Matthijs Douze}, \bibinfo{person}{Artem Babenko}, \bibinfo{person}{Dmitry Baranchuk}, \bibinfo{person}{Qi Chen}, \bibinfo{person}{Lucas Hosseini}, \bibinfo{person}{Ravishankar Krishnaswamny}, \bibinfo{person}{Gopal Srinivasa}, {et~al\mbox{.}}} \bibinfo{year}{2022}\natexlab{}.
\newblock \showarticletitle{Results of the NeurIPS’21 challenge on billion-scale approximate nearest neighbor search}. In \bibinfo{booktitle}{\emph{NeurIPS 2021 competitions and demonstrations track}}. \bibinfo{pages}{177--189}.
\newblock


\bibitem[Tan et~al\mbox{.}(2021)]%
        {tan2021norm}
\bibfield{author}{\bibinfo{person}{Shulong Tan}, \bibinfo{person}{Zhaozhuo Xu}, \bibinfo{person}{Weijie Zhao}, \bibinfo{person}{Hongliang Fei}, \bibinfo{person}{Zhixin Zhou}, {and} \bibinfo{person}{Ping Li}.} \bibinfo{year}{2021}\natexlab{}.
\newblock \showarticletitle{Norm adjusted proximity graph for fast inner product retrieval}. In \bibinfo{booktitle}{\emph{Proceedings of the 27th ACM SIGKDD Conference on Knowledge Discovery \& Data Mining}}. \bibinfo{pages}{1552--1560}.
\newblock


\bibitem[Tian et~al\mbox{.}(2023)]%
        {tian2023db}
\bibfield{author}{\bibinfo{person}{Yao Tian}, \bibinfo{person}{Xi Zhao}, {and} \bibinfo{person}{Xiaofang Zhou}.} \bibinfo{year}{2023}\natexlab{}.
\newblock \showarticletitle{DB-LSH 2.0: Locality-sensitive hashing with query-based dynamic bucketing}.
\newblock \bibinfo{journal}{\emph{IEEE Transactions on Knowledge and Data Engineering}} \bibinfo{volume}{36}, \bibinfo{number}{3} (\bibinfo{year}{2023}), \bibinfo{pages}{1000--1015}.
\newblock
\newblock
\shownote{{\underline{\color{blue}\href{https://doi.org/10.1109/TKDE.2023.3295831} {https://doi.org/10.1109/TKDE.2023.3295831}}}}.


\bibitem[Wang et~al\mbox{.}(2021)]%
        {wang2021comprehensive}
\bibfield{author}{\bibinfo{person}{Mengzhao Wang}, \bibinfo{person}{Xiaoliang Xu}, \bibinfo{person}{Qiang Yue}, {and} \bibinfo{person}{Yuxiang Wang}.} \bibinfo{year}{2021}\natexlab{}.
\newblock \showarticletitle{A comprehensive survey and experimental comparison of graph-based approximate nearest neighbor search}.
\newblock \bibinfo{journal}{\emph{arXiv preprint arXiv:2101.12631}} (\bibinfo{year}{2021}).
\newblock


\bibitem[Wei et~al\mbox{.}(2020)]%
        {wei2020analyticdb}
\bibfield{author}{\bibinfo{person}{Chuangxian Wei}, \bibinfo{person}{Bin Wu}, \bibinfo{person}{Sheng Wang}, \bibinfo{person}{Renjie Lou}, \bibinfo{person}{Chaoqun Zhan}, \bibinfo{person}{Feifei Li}, {and} \bibinfo{person}{Yuanzhe Cai}.} \bibinfo{year}{2020}\natexlab{}.
\newblock \showarticletitle{Analyticdb-v: A hybrid analytical engine towards query fusion for structured and unstructured data}.
\newblock \bibinfo{journal}{\emph{Proceedings of the VLDB Endowment}} \bibinfo{volume}{13}, \bibinfo{number}{12} (\bibinfo{year}{2020}), \bibinfo{pages}{3152--3165}.
\newblock


\bibitem[Wei et~al\mbox{.}(2024)]%
        {wei2024det}
\bibfield{author}{\bibinfo{person}{Jiuqi Wei}, \bibinfo{person}{Botao Peng}, \bibinfo{person}{Xiaodong Lee}, {and} \bibinfo{person}{Themis Palpanas}.} \bibinfo{year}{2024}\natexlab{}.
\newblock \showarticletitle{DET-LSH: A Locality-Sensitive Hashing Scheme with Dynamic Encoding Tree for Approximate Nearest Neighbor Search}.
\newblock \bibinfo{journal}{\emph{Proceedings of the VLDB Endowment}} \bibinfo{volume}{17}, \bibinfo{number}{9} (\bibinfo{year}{2024}), \bibinfo{pages}{2241--2254}.
\newblock
\newblock
\shownote{{\underline{\color{blue}\href{https://doi.org/10.14778/3665844.3665854} {https://doi.org/10.14778/3665844.3665854}}}}.


\bibitem[Woo et~al\mbox{.}(2023)]%
        {woo2023convnext}
\bibfield{author}{\bibinfo{person}{Sanghyun Woo}, \bibinfo{person}{Shoubhik Debnath}, \bibinfo{person}{Ronghang Hu}, \bibinfo{person}{Xinlei Chen}, \bibinfo{person}{Zhuang Liu}, \bibinfo{person}{In~So Kweon}, {and} \bibinfo{person}{Saining Xie}.} \bibinfo{year}{2023}\natexlab{}.
\newblock \showarticletitle{Convnext v2: Co-designing and scaling convnets with masked autoencoders}. In \bibinfo{booktitle}{\emph{Proceedings of the IEEE/CVF conference on computer vision and pattern recognition}}. \bibinfo{pages}{16133--16142}.
\newblock


\bibitem[Zhao et~al\mbox{.}(2023)]%
        {zhao2023fargo}
\bibfield{author}{\bibinfo{person}{Xi Zhao}, \bibinfo{person}{Bolong Zheng}, \bibinfo{person}{Xiaomeng Yi}, \bibinfo{person}{Xiaofan Luan}, \bibinfo{person}{Charles Xie}, \bibinfo{person}{Xiaofang Zhou}, {and} \bibinfo{person}{Christian~S Jensen}.} \bibinfo{year}{2023}\natexlab{}.
\newblock \showarticletitle{FARGO: Fast maximum inner product search via global multi-probing}.
\newblock \bibinfo{journal}{\emph{Proceedings of the VLDB Endowment}} \bibinfo{volume}{16}, \bibinfo{number}{5} (\bibinfo{year}{2023}), \bibinfo{pages}{1100--1112}.
\newblock


\bibitem[Zhou et~al\mbox{.}(2019)]%
        {zhou2019mobius}
\bibfield{author}{\bibinfo{person}{Zhixin Zhou}, \bibinfo{person}{Shulong Tan}, \bibinfo{person}{Zhaozhuo Xu}, {and} \bibinfo{person}{Ping Li}.} \bibinfo{year}{2019}\natexlab{}.
\newblock \showarticletitle{M{\"o}bius transformation for fast inner product search on graph}.
\newblock \bibinfo{journal}{\emph{Advances in Neural Information Processing Systems}}  \bibinfo{volume}{32} (\bibinfo{year}{2019}).
\newblock


\end{thebibliography}
\end{document}